\documentclass[nonacm,screen,acmsmall]{acmart}
\def\setextended{}

\usepackage[shortlabels,inline]{enumitem}
\usepackage{thmtools}
\usepackage{mathtools}
\usepackage{bm}
\usepackage{multirow}
\usepackage{tikz}
\usepackage{subcaption}
\usepackage{cleveref}

\usetikzlibrary{arrows.meta}
\usetikzlibrary{arrows}
\usetikzlibrary{automata}
\usetikzlibrary{shapes.geometric}

\tikzset{round/.style={circle,draw,minimum size=18pt,inner sep=0pt}}
\tikzset{double round/.style={
		round, double, double distance=2pt,
		minimum size=16pt
}}

\newif\ifcomments
\ifdef{\setcomments}{\commentstrue}{\commentsfalse}

\newif\ifproofs
\ifdef{\setproofs}{\proofstrue}{\proofsfalse}

\newif\ifextended
\ifdef{\setextended}{\extendedtrue}{\extendedfalse}

\ifcomments
\newcommand{\sharon}[1]{\textcolor{magenta}{\textbf{SH:} #1}}
\newcommand{\sharonnew}[1]{\textcolor{purple}{\textbf{SH:} #1}}
\newcommand{\neta}[1]{\textcolor{cyan}{\textbf{NE:} #1}}
\newcommand{\adi}[1]{{\color{orange}AM: #1}}
\else
\newcommand{\sharon}[1]{\ifdim\lastskip>0pt\ignorespaces\fi}
\newcommand{\sharonnew}[1]{\ifdim\lastskip>0pt\ignorespaces\fi}
\newcommand{\neta}[1]{\ifdim\lastskip>0pt\ignorespaces\fi}
\newcommand{\adi}[1]{\ifdim\lastskip>0pt\ignorespaces\fi}
\fi

\newcommand{\acronym}[1]{{\ifmmode \mathchoice {\textup{\MakeUppercase{#1}}}{\textup{\MakeUppercase{#1}}}{\textup{\textsc{\MakeLowercase{#1}}}}{\textup{\textsc{\MakeLowercase{#1}}}}\else \MakeUppercase{#1}\fi }}
\newcommand{\mprog}[1]{\ensuremath{{\texttt{\footnotesize{#1}}}}}

\newcommand{\Dom}{\mathcal{D}}
\newcommand{\Interp}{\mathcal{I}}
\newcommand{\Vars}{\mathcal{X}}
\newcommand{\Power}{\mathcal{P}}

\newcommand{\Z}{\mathbb{Z}}

\newcommand{\Logic}{\mathcal{L}}
\newcommand{\Bound}{\mathcal{B}}
\newcommand{\Explic}[1]{\mathcal{E}\parens{#1}}
\newcommand{\Models}{\mathcal{M}}

\newcommand{\sort}{\bm{s}}
\newcommand{\fgSort}{\sort^*}

\newcommand{\prog}[1]{\texttt{#1}}
\newcommand{\lseg}{\prog{lseg}}

\newcommand{\slist}{\prog{slist}}
\newcommand{\keys}{\prog{keys}}
\newcommand{\key}{\prog{key}}
\newcommand{\covered}{\prog{covered}}
\newcommand{\conflict}{\prog{conflict}}
\newcommand{\unmatchTop}{\prog{unmatch\_top}}
\newcommand{\unmatchRight}{\prog{unmatch\_right}}

\newcommand{\lia}{\acronym{lia}}
\newcommand{\qfLia}{\textrm{QF}_\lia}
\newcommand{\intSort}{\ensuremath{\textbf{\mprog{int}}}}
\newcommand{\intTheory}{\lia}
\newcommand{\intModel}{M^\intTheory}
\newcommand{\intVocab}{\Sigma^\intTheory}
\newcommand{\intAxioms}{\Gamma^\intTheory}

\newcommand{\FOEncoding}{\mathcal{F}}
\newcommand{\eVar}{x^*}

\newcommand{\defeq}{\triangleq}
\newcommand{\bigland}{\mathop{\bigwedge}}
\newcommand{\biglor}{\mathop{\bigvee}}
\newcommand{\liff}{\mathrel{\leftrightarrow}}
\newcommand{\Substitute}[2]{#2 / #1}
\newcommand{\SingSubstitute}[2]{\brackets{\Substitute{#1}{#2}}}
\newcommand{\entails}{\vdash}
\DeclareDocumentCommand{\sysentails}{O{\sys}}{\vdash^{#1}}
\newcommand{\Quantifier}{\mathcal{Q}}

\newcommand{\wsl}{\acronym{wsl}}
\newcommand{\seplog}{\acronym{sl}}
\newcommand{\lfp}{\acronym{lfp}}
\newcommand{\fp}{\acronym{fp}}
\newcommand{\fo}{\acronym{fo}}
\newcommand{\fol}{\acronym{fol}}

\newcommand{\eprsl}{\acronym{edh}}
\newcommand{\eprwsl}{\acronym{edh(wsl)}}

\newcommand{\fg}{\acronym{sl}}
\newcommand{\bg}{\ensuremath{\mathcal{T}}}

\DeclareMathOperator{\sk}{sk}

\DeclareMathOperator{\FreeVars}{FV}
\DeclareMathOperator{\true}{true}
\DeclareMathOperator{\false}{false}

\DeclareMathOperator{\fold}{fold}
\DeclareMathOperator{\unfold}{unfold}

\DeclarePairedDelimiter{\pars}{\lparen}{\rparen}
\newcommand{\parens}[1]{\pars*{#1}}
\DeclarePairedDelimiterX{\curlybracs}[1]{\lbrace}{\rbrace}{#1}
\newcommand{\braces}[1]{\curlybracs*{#1}}
\DeclarePairedDelimiter{\angles}{\langle}{\rangle}
\newcommand{\tuple}[1]{\angles*{#1}}
\DeclarePairedDelimiter{\bracs}{\lbrack}{\rbrack}
\newcommand{\brackets}[1]{\bracs*{#1}}
\DeclarePairedDelimiter{\bbracs}{\llbracket}{\rrbracket}
\newcommand{\bbrackets}[1]{\bbracs*{#1}}
\DeclarePairedDelimiterX{\midbracs}[2]{\lbrace}{\rbrace}{\,#1\;\delimsize\vert\;#2\,}
\newcommand{\midbraces}[2]{\midbracs*{#1}{#2}}

\renewcommand{\Vec}[1]{\bm{#1}}
\newcommand{\tdots}{\times\dots\times}

\newcommand{\partto}{\rightharpoonup}
\newcommand{\spacedMid}{\ensuremath{\,\mid\,}}

\newcommand{\points}[1]{\mapsto\tuple{#1}}
\newcommand{\pointsi}{\mapsto}
\newcommand{\nil}{\ensuremath{nil}}
\newcommand{\Null}{\mprog{null}}
\newcommand{\emp}{\ensuremath{emp}}
\newcommand{\sys}{\Phi}

\newcommand{\tightarray}{\arraycolsep=1.6pt}

\newcommand{\benchmark}[1]{\textbf{\textsc{#1}}}
\newcommand{\subbench}[1]{\textit{#1}}
\newcommand{\silentSmallskip}{\par\addvspace{0.4\baselineskip}}
\newcommand{\mypara}[1]{\silentSmallskip{}\noindent\emph{\textbf{#1}.\ }}
\newcommand{\smallpara}[1]{\silentSmallskip{}\textit{#1.}}

\newcommand{\zzz}{Z3}
\newcommand{\fest}{FEST}
\newcommand{\songbird}{Songbird}
\newcommand{\sls}{SLS}

\newcommand{\around}[1]{${\sim}#1$}

\AtEndPreamble{
	\newtheorem{claim}[theorem]{Claim}
	\newtheorem{conclusion}[theorem]{Conclusion}
	\theoremstyle{definition}
	
	\newtheorem*{note}{Note}
	\newtheorem*{notation}{Notation}
	\newtheorem*{remark}{Remark}
	
	\crefname{lemma}{lemma}{lemmas}
	\crefname{claim}{claim}{claims}
	\crefname{figure}{fig.}{figs.}
	\crefname{theorem}{thm.}{thms.}
	\crefname{definition}{def.}{defs.}
	\crefname{equation}{eq.}{eqs.}
}

\newcommand{\Proofs}{}

\NewDocumentCommand{\HereProof}{m+m}{\begin{proof}#2
	\end{proof}}

\NewDocumentCommand{\ProofOf}{m+m}{\begin{proof}[Proof of \Cref{#1}]\phantomsection
		\label{proof:#1}
		#2
	\end{proof}}

\ifproofs
\NewDocumentCommand{\Proof}{m+m}{\HereProof{#1}{#2}}
\else
\NewDocumentCommand{\Proof}{m+m}{\neta{proof deferred to \Cref{sec:proofs}: \hyperref[proof:#1]{proof}}
	\toks0=\expandafter{\Proofs}\toks2=\expandafter{\ProofOf{#1}{#2}}\xdef\Proofs{\the\toks0 \the\toks2}}
\fi  
\title[Separating the Wheat from the Chaff]{Separating the Wheat from the Chaff:
	Understanding (In-)Completeness of 
	Proof Mechanisms for
	Separation Logic 
	with Inductive Definitions
}

\begin{document}

\setcopyright{cc}
\setcctype{by}
\acmDOI{10.1145/3776671}
\acmYear{2026}
\acmJournal{PACMPL}
\acmVolume{10}
\acmNumber{POPL}
\acmArticle{29}
\acmMonth{1}
\received{2025-07-10}
\received[accepted]{2025-11-06} 
\begin{abstract}
    \neta{this is a small comment to make sure I compiled without comments}For over two decades Separation Logic has enjoyed
its unique position as arguably the most popular framework
for reasoning about heap-manipulating programs,
as well as reasoning about shared resources and permissions.
Separation Logic is often extended to include
inductively-defined predicates,
interpreted as least fixpoints,
to form what is known as Separation Logic with Inductive Definitions (SLID).
These inductive predicates are
used to describe unbounded data-structures in the heap
and to verify key properties thereof.
Many theoretical and practical advances 
have been made in developing automated proof mechanisms
for SLID,
but by their very nature these mechanisms are imperfect,
and a deeper understanding of their failures 
is desired.
As expressive as Separation Logic is, it is not surprising
that it is incomplete:
there is no procedure that will provide a proof
for all valid entailments in Separation Logic.
In fact, at its very core, Separation Logic contains several
sources of incompleteness that defy automated reasoning.

In this paper we study these sources of incompleteness
and how they relate to failures of proof mechanisms of SLID.
We contextualize SLID within a larger, relaxed logic,
that we call Weak Separation Logic (WSL).
We prove that unlike SLID, WSL enjoys completeness 
for a non-trivial
fragment of quantified entailments with background theories and 
inductive definitions,
via a reduction to first-order logic (FOL).
Moreover, we show that the ubiquitous fold/unfold proof mechanism,
which is unsurprisingly incomplete for SLID,
does constitute a sound and complete proof mechanism of 
WSL,
for theory-free, quantifier-free entailments with inductive definitions.
In some sense, this shows that WSL is the natural logic
of such proof mechanisms.
Through this contextualization of SLID within WSL,
we understand proof failures as stemming from
rogue, nonstandard models,
that exist within the class of models considered by WSL,
but do not adhere to the stricter requirements of SLID.
These rogue models are typically infinite,
and we use the
recently proposed formalism of symbolic structures
to represent and automatically find them.

We present a prototype tool that implements the encoding of 
WSL to FOL and test it on an existing benchmark,
which contains over 700 
quantified entailment problems with inductive definitions,
a third of which also contain background theories.
Our tool is able to find counter-models
to many of the examples,
and we provide a partial taxonomy of the rogue models,
shedding some light on real-world proof failures.
 \end{abstract}

\author{Neta Elad}
\orcid{0000-0002-5503-5791}
\affiliation{\institution{Tel Aviv University}
  \city{Tel Aviv}
  \country{Israel}
}
\email{netaelad@mail.tau.ac.il}

\author{Adithya Murali}
\orcid{0000-0002-6311-1467}
\affiliation{\institution{University of Wisconsin-Madison}
  \city{Madison}
  \state{WI}
  \country{USA}
}
\email{adithyamurali@cs.wisc.edu}

\author{Sharon Shoham}
\orcid{0000-0002-7226-3526}
\affiliation{\institution{Tel Aviv University}
  \city{Tel Aviv}
  \country{Israel}
}
\email{sharon.shoham@gmail.com} 

 
\maketitle

\neta{consistent terminology and todos:
	\begin{itemize}
		\item in Dryad, heaplet refers to the heap locations + mapping
		(not just the support), 
		and what we call heaplets they simply call locations (denoted $R$).
		\item change entailment symbol to $\models$,
			or more specifically $\models_\Logic$ where
			$\Logic \in {SL, WSL}$.
		\item
		$P$ for foreground/recursive predicates, 
		$Q$ for background,
		$P_\fo$ for FO relations (domain of $P$),
		$P_\eta$ for heaplets.
		\item What's the relation between QF, conjunctive formulas (in SL)
		and so-called ``symbolic heap'' formulas? Are they the same?
		Symbolic heap = conjunctive formulas?
		\item 
		We need to be clear that we are defining SL without magic wand;
		The BSR(SL) shows undecidability, which seems contradictory to our results,
		but they allow magic wand.
		\item 
		theory predicates vs. inductive predicates,
		theory vocabulary vs. SL vocabulary,
		$\fgSort$ is the location sort. 
		\item 
		\lfp{} struct for or \lfp{} struct of.
		\item the phrase ``background theory'' is used in SL papers,
		so we don't need to avoid it.
		\item 
		radical suggestion: use SLID (maybe also WSLID) for the name of the logic,
		but 
		$\vdash_{\mathcal{S}}, \vdash_{\mathcal{W}}$ for notation.
		otherwise we need to make it very clear (in the overview? first technical section)
		that we refer to SLID by SL.
		\item heap vs. heaplet
		\item 
		unique heap for base (inductive predicate),
		determined heap for fragment
		\item $\nil$ (the constant) vs. $\Null$ (the element): 
		maybe the distinction is unimportant, or even confusing
		\item put ``wheat'' and ``chaff'' somewhere 
		in the body/overview of the paper
		\item inside a sentence separation logic or with capitals
		\item consistent order: theory-free, quantifier-free 
		or quantifier-free, theory-free?
		no comma?
		\item use $\Delta$ for finite set of assumption?
		might also differentiate it from $\Gamma^\bg$.
	\end{itemize}
}

\section{Introduction}
\label{sec:intro}

Automated techniques for program verification have made incredible advances in the last decade owing to the development of effective algorithms for checking logical validity and entailment. The most notable examples of this are SMT solvers~\cite{Z3,cvc4,cvc5}, which implement decision procedures for validity of various combinations of quantifier-free decidable theories~\cite{nelson-oppen1979}. The development of theory and algorithms for decidable combinations of theories has been a critical foundation for the many successes of automated verification.

However, from the perspective of practice there remain significant gaps in our foundations. Practical program verification frequently requires the use of expressive logics for which decision procedures cannot exist; in fact, most of them are incomplete! Plainly, there cannot exist sound proof systems or validity procedures that can prove every valid formula. Furthermore, there are no canonical approaches for automating validity for such logics, and the theoretical limitations of existing automation techniques are not well-understood.

This lack of theoretical grounding comes at a practical cost. Verification tools implement heuristic approaches for reasoning with incomplete logics that can fail to prove valid formulas without clear reason. This is a particularly frustrating form of failure, since in practice it can be hard to distinguish plainly invalid formulas from valid formulas on which particular heuristics meet their limitations. Additionally, even savvy users who are willing to work through the failures can find systematic approaches for doing so lacking. This absence of `diagnosability' breeds further frustration.

Thus, a foundational understanding of the incompleteness of validity/entailment checking mechanisms for various expressive logics is sorely needed.

\subsection*{Separation Logic with Inductive Definitions and Background Theories}
Separation Logic (\seplog{}) is arguably the most popular logic for specifying and reasoning about heap-manipulating programs~\cite{demri15,seplogicprimer,ohearn01,reynolds02}. Formulas are interpreted over a store $s$ and a finite heap $h$, and can express a combination of \emph{pure} properties that only involve the store as well as \emph{spatial} properties that describe heaplets (portions of a heap) and separations between them. In practice, separation logic is augmented with inductively defined predicates to describe unbounded structures on the heap and their properties, e.g., linked lists, trees, sorted-ness, length/height, set of keys stored in the structure, etc.
These predicates are interpreted as the least fixpoint of their definitions. The logic is also usually presented with a combination of background theories for reasoning about data values. We refer to this logic as SLID (Separation Logic with Inductive Definitions).

Unsurprisingly, SLID is incomplete.
Formally, checking the validity of an entailment
between SLID formulas
is not recursively enumerable.
There are many sources of incompleteness in SLID:
\begin{enumerate*}[(a)]
	\item
	the magic wand,\item
	the consideration of validity only over finite heaplets,\item
	least fixpoints in the inductive definitions,\item
	combinations with intended models of background sorts used for data reasoning,
	and
	\item
	implicit second-order quantification over heaplets in the separating conjunction.\footnote{All of these sources except for the last one are
well known causes of incompleteness in logics. Finite model
theory is known to be incomplete for First-Order Logic,
least fixpoints can define arbitrary computations, and
logics that combine, say integers and uninterpreted
functions can already express multiplication which yields
incompleteness. We discuss these arguments elaborately in
the context of separation logic in
\ifextended
\Cref{sec:background}.
\else
the extended version of the paper~\cite{wheat-chaff-extended}.
\fi
}
\end{enumerate*}

Among these sources, it is widely agreed  that the magic wand is difficult to reason with (it is essentially a form of second-order quantification~\cite{magic-wand-complexity}), and in fact many automated tools do not support it~\cite{slcomp}. We therefore consider the automation of separation logics without the magic wand, and focus on the impact of the four latter sources of incompleteness.

\mypara{The Research Questions}
There are several approaches to dealing with incompleteness in the literature. Some works focus on identifying decidable fragments of SLID, while others design new heap logics that enjoy certain completeness properties~\cite{BerdineCalcagnoOHearn2004,BerdineCalcagnoOHearn2005,twbcade13,focomplete-heap-logics}.

In this work we seek a different pragmatic end: \emph{Why} does existing automation for separation logic fail to be predictable? In other words, which among the many sources of incompleteness cause proof mechanisms to fail? What are the formal connections between the sources of incompleteness and the failure of proof for specific entailments, and more importantly, what does such a theory say about how to overcome proof failures? 

We focus our investigation on one of the most popular techniques for automating validity checking of entailments in SLID:
fold/unfold-based reasoning. Variants of this technique are implemented in several automated verifiers and entailment checkers for SLID~\cite{songbird,vipertool,focomplete-heap-logics,qiu13,ChinDavidNguyen2007}.

\mypara{Fold/Unfold Proof Mechanisms} The core idea in fold/unfold-based reasoning is the following: given an inductively defined predicate $P(\Vec{x})$ with definition $\rho(\Vec{x})$, these mechanisms \emph{unfold} the body of the definition on predicate terms occurring in the entailment. More precisely, when a term $P(\Vec{t})$ occurs in an entailment, they replace it with $\rho(\Vec{t})$ and treat any further occurrences of $P$ as an uninterpreted predicate.
Similarly, these mechanisms may also \emph{fold} a definition,
replacing $\rho(\Vec t)$ with $P(\Vec t)$.
They then try to prove the resulting entailment valid in SL (without inductive definitions) using a complete proof mechanism. This is clearly an overapproximation, in the sense that proving the unrolled entailment valid implies the validity of the original entailment. Of course, the converse may not hold in general.

Since SLID is incomplete, all sound proof mechanisms
--- including fold/unfold mechanisms ---
cannot prove every valid entailment. However, the limitations are not merely theoretical. These mechanisms do in fact fail to prove valid entailments that arise in practical program verification~\cite{vipercav24}, and users can find themselves stuck not knowing what to do next. The lack of theoretical foundations to understand or deal with these failures is a key obstacle towards making practical and predictable automation for separation logic possible.

\begin{quote}
\emph{In this work we develop foundations for understanding the (in-)completeness of fold/unfold mechanisms for checking entailments in Separation Logic with Inductive Definitions.}
\end{quote}

\subsection*{Logical Characterizations of (In-)Completeness and Rogue Models}
We address the aforementioned research questions by developing
\emph{logical characterizations}
of the (in-)completeness of fold/unfold proof mechanisms
for separation logic.
Our approach is inspired by recent works~\cite{loding18,fluid23}
which pursue the following idea:
Let $\mathcal{L}_1$ be a logic whose formulas
are interpreted over a class of models $\mathcal{M}_1$,
and let $\mathcal{S}$ be a proof mechanism
which is incomplete for $\mathcal{L}_1$,
i.e., it cannot prove every entailment
that is valid in $\mathcal{L}_1$.
The (in-)completeness of $\mathcal{S}$
is characterized by defining a logic $\mathcal{L}_2$
whose syntax (and satisfaction relation)
is the same as that of $\mathcal{L}_1$,
but where valid formulas (or entailments) must hold on a class of models $\mathcal{M}_2$ which is larger than $\mathcal{M}_1$. The technique then involves showing that $\mathcal{S}$ is a \textbf{sound proof mechanism} for entailments in $\mathcal{L}_2$. This means that the proving power of $\mathcal{S}$ is bounded by a \emph{weaker} logic. Note that if an entailment is valid in $\mathcal{L}_2$, then it is certainly valid in $\mathcal{L}_1$, since $\mathcal{L}_2$ is interpreted over a larger class of models. However, the converse need not hold. Importantly, this means that entailments that hold in $\mathcal{L}_1$ but not in $\mathcal{L}_2$ \emph{cannot} be proven using $\mathcal{S}$.
This helps explain why $\mathcal{S}$ can fail to prove some entailments that are valid in $\mathcal{L}_1$: it is because these entailments may actually be invalid in $\mathcal{L}_2$, which is in some sense the ``true logic'' over which $\mathcal{S}$ operates.

The development of a logical characterization also yields an interesting way to witness the limitations of $\mathcal{S}$ in proving specific entailments. Suppose that an entailment $\alpha \models \beta$ is valid in $\mathcal{L}_1$ but invalid in $\mathcal{L}_2$. This means that there is a model $M$ belonging to the class $\mathcal{M}_2$ such that $M$ satisfies $\alpha$ but not $\beta$. Note that $M$ cannot belong to $\mathcal{M}_1$, since the entailment is valid in $\mathcal{L}_1$. Such a model $M$ is referred to as a \textbf{rogue model}. The term rogue model refers to the fact that a user using $\mathcal{S}$ to prove entailments in $\mathcal{L}_1$ thinks about the validity of formulas over the intended or \emph{standard} class of models $\mathcal{M}_1$, but their attempt is foiled by the \emph{rogue} model $M$ that lies outside this class. Additionally, if $\mathcal{S}$ is \textbf{complete} for $\mathcal{L}_2$, then it follows that \emph{every} failure of $\mathcal{S}$ to prove an entailment valid in $\mathcal{L}_1$ can be traced to the existence of a rogue model. A logical characterization of the completeness thus helps not only understand and compare the relative completeness or incompleteness of proof mechanisms,
but also yields a way to connect the incompleteness of $\mathcal{S}$ to its failure to prove specific entailments via rogue models.

\silentSmallskip{}
We note here that the works we draw inspiration from do not explicitly develop the idea of a logical characterization of completeness. We distill the approach employed in these works into the above insight and apply the framework to separation logic.
This is nontrivial, and is also the first such
investigation of separation logic
as prior works only studied first-order logics.

\subsection*{Main Contributions}

The central insight of this work is that the (in-)completeness of fold/unfold mechanisms can be understood through the lens of a new separation logic that is constructed from standard SLID by ablating the various sources of incompleteness. By defining this weaker separation logic and studying its models, we are able to predict the failures of fold/unfold mechanisms on SLID entailments, connect the failures to different sources of incompleteness, and formulate systematic recommendations for overcoming them. These insights are the result of the following contributions:

\mypara{1. A Weak Separation Logic (Section~\ref{sec:wsl})} Our first contribution is the development of a new separation logic to explore the influence of the various sources of incompleteness in standard SLID. We do this by systematically relaxing the various design decisions that lead to incompleteness in SLID. We remove the influence of least fixpoints by allowing interpretations for inductively defined predicate symbols to conform to any fixpoint of the definition. We remove the incompleteness arising from considering only finite heaplets by defining satisfaction of formulas over both finite and infinite heaplets and defining validity to mean satisfaction over both finite and infinite heaplets. We also ablate the incompleteness arising from requiring standard models of background theories
by considering instead the class of all models of the theories
\neta{changed
(e.g., all models of Presburger Arithmetic
as opposed to just the standard number line model).}
(e.g., allowing nonstandard models of Presburger Arithmetic).
We call the resulting logic
\textbf{Weak Separation Logic} (\wsl{}).

While it may be common to tacitly consider such relaxations when developing proof mechanisms for logics, our key insight is that we can perform the process in reverse: namely that we can characterize fold/unfold mechanisms by pursuing certain relaxations. This is rather subtle. In fact, while one may expect the above (perhaps usual) relaxations to be sufficient to eliminate all incompleteness, we show that performing any combination of the above relaxations ---
or indeed all of them --- is not sufficient 
\ifextended
(see~\Cref{sec:background}).
\else
(see details in the extended version 
of the paper~\cite{wheat-chaff-extended}).
\fi

The remaining source of incompleteness of \wsl{}
is the second-order quantification
introduced by the separating conjunction:
an SL formula $\alpha * \beta$
holds on a heaplet $h$
if there exists a partitioning of $h$
such that one partition satisfies $\alpha$
and the other satisfies $\beta$.
Similarly, showing that the formula does not hold
requires an argument for all possible partitionings.
This weak form of second-order quantification turns out to be
sufficient to yield incompleteness 
\ifextended
(see~\Cref{sec:background}).
\else
(see details in the extended version
of the paper~\cite{wheat-chaff-extended}).
\fi
Further, it poses challenges for the relaxation
of finite heaplets to finite and infinite heaplets,
as infinite heaplets may have infinitely many partitionings.
\neta{no exacerbate:
The relaxations exacerbate the difficulty of dealing
with the separating conjunction since heaplets may now be infinite,
and may have infinitely many partitionings.
}

We tackle this by first observing that many inductive definitions
that arise in practice are well-behaved in a certain sense:
in any particular global heap,
there can be at most one heaplet that satisfies an inductive predicate.
This is natural;
after all, formulas are meant to capture specific portions of memory.
We then define a fragment of \wsl{},
called the Effectively Determined-Heap (\eprsl{}) fragment,
where a weaker form of this property is lifted to all formulas,
ensuring that
only finitely many heaplets must be considered for each formula.
The combination
of the relaxed semantics of \wsl{}
and the \eprsl{} restrictions
allow us to establish completeness through a
validity-preserving translation into first-order logic,
generalizing similar existing encodings (e.g.,~\cite{separation-predicates}).

The \fo{} translation is a significant contribution because
it allows us to transfer established theoretical techniques
for analyzing first-order logic
and apply them to separation logics.
In fact, it immediately yields that the \eprsl{} fragment of \wsl{}
admits complete validity checking for entailments!
\neta{changed: This is a nontrivial result }
This is nontrivial
since the fragment is quite rich,
and in particular allows formulas
\neta{changed:
to include quantification
over heap locations (with certain restrictions). }
to include some quantification over heap locations.
In fact, it is undecidable.
The connection between \wsl{} and \fol{}
is used crucially to analyze the completeness
of fold/unfold mechanisms.

\mypara{2. Soundness and Completeness of
Fold/Unfold Reasoning for WSL (Section~\ref{sec:unfolding})} Our second contribution is an analysis of the proving power of fold/unfold mechanisms for WSL. We formally model a large class of fold/unfold mechanisms and show these mechanisms are sound for WSL.
This is an important result, because it establishes an upper bound on the power of the proof mechanisms. Entailments that are invalid in WSL cannot be proven by fold/unfold mechanisms, and therefore fold/unfold mechanisms are bound to fail for such entailments even when they are valid in SLID.

Completeness of fold/unfold reasoning for WSL
is much more subtle to pose as it depends on the specific unfolding strategies,
and variants can differ in proving power.
Following the intuitive connection between
fold/unfold operations and quantifier instantiation
(employed in,
e.g.,~\cite{sl-and-abstraction,abstract-existential-types,hosl-abstraction}),
our broad insight for proving completeness
is that we can formalize this connection
and establish a natural correspondence
between fold/unfold transformations in \wsl{}
and quantifier instantiations in \fol{}.
We utilize this to show that
validity of entailments in quantifier-free theory-free \wsl{}
can always be proved by finitely many
fold/unfold steps,
and, thus,
mechanisms that fold and unfold
definitions on all terms are
complete for this fragment of
\wsl{}. This result is a culmination of several developments,
including the fact that WSL entailments admit a
validity-preserving translation to FOL, the correspondence
between fold/unfold axioms and quantifier instantiations,
and properties of first-order logic including Herbrand's
theorem and compactness. This is another instance where our
contributions offer a way to use mathematical tools from the
FOL literature in the context of separation logics.

Importantly, our soundness and completeness results show that WSL serves as a logical characterization of the completeness of fold/unfold mechanisms. Our results also imply that the failure of fold/unfold mechanisms to prove entailments in SLID without theories and quantifiers can be \emph{entirely} explained by WSL. We believe that the evidence strongly points to a similar result for SLID with theories and (restricted) quantification as well, where fold/unfold applications are combined with other quantifier instantiations to handle first-order quantifiers and theories.

\mypara{3. Symbolic Representation and Synthesis of Rogue
Models (Section~\ref{sec:rogue-models})} We then turn to the
connection between WSL and the failure of fold/unfold
mechanisms on specific entailments. Our completeness results
show that when fold/unfold mechanisms fail to prove an
entailment in SLID, we can look for a rogue model on which
the entailment, interpreted in WSL, does not hold. To find
such rogue models we adopt the mathematical framework of
symbolic structures developed in recent
work~\cite{infinite-needle}, which defines a class of
infinite models for first-order logic that admit finitary
symbolic representations and efficient synthesis for
refutation.
Once again, we utilize the validity-preserving
(and model-preserving) translation of \wsl{} entailments to \fol{}.

We extend the theoretical developments in the prior work to utilize the algorithmic techniques in the context of separation logic.
In particular, we extend symbolic structures
with the background theory of the integers,
synthesizing rogue models that are extensions of the standard model of the integers
for practical SLID entailments.
We also identify a large class of inductive definitions
called \emph{heap-reducing} definitions for which rogue
models, if they exist, must be infinite. To our knowledge,
almost every inductive definition encountered in practice
falls into this class. This substantiates our approach of
representing rogue models using a framework for expressing
infinite structures.

\mypara{4. Empirical Evaluation of Rogue Models for SLID Entailments (Section~\ref{sec:eval})} We implement a tool for checking validity of SLID entailments under WSL semantics via our translation to FOL.
The tool dually attempts to
\begin{enumerate*}[(a)]
	\item
	prove validity of entailments under WSL semantics
	(which would of course show validity in SLID), and
	\item
	synthesize rogue models which would show that it is not valid in WSL semantics
	(and could be nontrivial to prove in SLID).
  \end{enumerate*}
 Our synthesis algorithm builds on the model-finding procedure
 developed in~\cite{infinite-needle},
 fine-tuning an abstract template to represent
 rogue models of practical SLID entailments
(extended to our setting with the theory of the integers
 as described above).
We evaluate our tool on a suite of over 700 entailments established in prior work~\cite{sls} and show that it is effective at proving entailments as well as finding rogue models. We also analyze the rogue models qualitatively and provide a partial taxonomy. To the best of our knowledge, this is one of only two systematic studies of rogue models performed at scale~\cite{BlanchetteLPAR2010}, and the only one on rogue models for heap logics.

\ifproofs\else
\ifextended
Proofs for the various claims and theorems throughout the paper
are deferred to \Cref{sec:proofs}.
\else
Proofs for the various claims and theorems throughout the paper
are available in the extended version of the paper~\cite{wheat-chaff-extended}.
\fi
\fi 

\section{Overview}
\label{sec:overview}

In this section we illustrate fold/unfold reasoning
for Separation Logic with Inductive Definitions~(SLID)
\neta{use SLID in overview}
and study its failures.
We relate proof failures to a semantic gap
between the intended semantics
and the actual, relaxed semantics that fold/unfold captures.
We demonstrate how rogue models,
which arise from the semantic gap,
can help understand the root cause of proof failure,
and guide the user towards the missing parts of the proof.

\subsection{Fold/Unfold Proof Mechanisms}
Consider the following inductive definition
of an acyclic list segment between two locations $x$ and $y$,
which starts from the base case where $x=y$
and extends the segment from the left,
building on a segment
between a location $u$ that $x$ points to
and $y$.
In the definition, the separating conjunction $*$ is used
to require that $x$ is
not in the set of locations
in the list segment between $u$ and $y$.
{\smaller\begin{equation}
	\label{eq:lseg-sys}
	\lseg(x,y) \coloneq x=y \lor
		\parens{x \neq y *
			\exists u. x\pointsi u * \lseg(u,y)
		}
\end{equation}}

For locations $x$ and $y$ in the heap,
$\lseg(x,y)$ characterizes
the set of locations
in the list segment between $x$ and $y$,
or, in other words,
the unique heaplet of $\lseg(x,y)$.
The definition in \Cref{eq:lseg-sys}
demonstrates a typical property of inductive definitions,
where each tuple of locations that satisfy an inductive predicate
has a unique, determined heaplet.
In fact, this matches a core design goal of separation logic,
where formulas are used to characterize heaplets.
Accordingly,
we only consider inductive definitions
that induce determined heaplets.
(As explained in \Cref{sec:intro},
this restriction also turns out to be paramount
for avoiding the incompleteness of SLID.)

One simple property of a list segment is that for any two distinct locations
$a, b$ that form a list segment,
there exists a location $v$ that $a$ points to,
such that $v$ and $b$ form a list segment.
This is written in separation logic as the following entailment.
{\smaller\begin{equation}
	\label{eq:lseg-valid}
	a\neq b * \lseg(a,b) \entails \exists v. a\pointsi v * \lseg(v,b)
\end{equation}}
This entailment can be proved by unfolding
$\lseg(a,b)$ into its definition,
resulting in
{\smaller\[
a\neq b * \parens{ a = b \lor
	\parens{a \neq b * \exists u. a\pointsi u * \lseg(u,b) }
}
\entails \exists v. a\pointsi v * \lseg(v,b),
\]}
which is clearly valid,
regardless of the interpretation of $\lseg$.
This example shows a successful application of a fold/unfold proof,
which more generally may unfold any instance of an inductive predicate
into its definition,
or, conversely, fold its definition into an application of the inductive predicate.
For example, the entailment
{\smaller\begin{equation}
	\label{eq:lseg-fold}
	a \neq b * a \pointsi{c} * \lseg(c, b) \entails \lseg(a,b)
\end{equation}}
can be proved by folding the antecedent into $\lseg(a,b)$.

Unfortunately, fold/unfold reasoning
may also fail to prove a valid entailment.
For example, consider the following entailment,
which states that a
list segment
can be extended from the right, i.e.,
obtaining a list segment between locations $a$ and $b$
from a list segment between $a$ and some location $c$
that points to $b$
which is not yet in the list
(ensured by a separating conjunction applied to $b \pointsi \nil$)
{\smaller\begin{equation}
	\label{eq:lseg-ind}
	\lseg(a,c) * c \pointsi b * b \pointsi \nil \entails \lseg(a,b) * b \pointsi \nil
\end{equation}}

This entailment is also valid, but cannot be proved
with any number of fold/unfold steps.
Intuitively,
the syntactic shape of the definition of $\lseg$
restricts fold/unfold reasoning to the left side of a segment,
but here
validity hinges on the points-to relation
$c \pointsi b$
between locations on the right side of segments.
Therefore, no matter how many fold/unfold steps are performed,
we will never be able to derive $\lseg(x,b)$
in conjunction with $a \pointsi^* x$,
which is required for ultimately deriving $\lseg(a,b)$.
Beyond this technical explanation,
the failure to prove \Cref{eq:lseg-ind}
can be understood on a deeper level
as a mismatch between
the intended semantics
and the semantics captured by fold/unfold.
In the intended semantics,
interpretations of \lseg{}
are least fixpoints,
hence all list segments are finite,
but fold/unfold proofs cannot distinguish least fixpoints from other fixpoints,
which also include infinite list segments.
Unfortunately, when
potentially infinite fixpoint interpretations of list segments
are considered,
the symmetry between the left and right sides of a segment,
which exists in finite segments,
is broken.
As a result,
list segments constructed by adding locations on the left
(as defined by $\lseg$)
cannot necessarily be constructed by adding locations on the right.
Thus, the entailment in \Cref{eq:lseg-ind}
is not valid over such interpretations.
As we will see in the sequel,
this abstract train of thought can be reified.

\subsection{Weak Separation Logic}
As illustrated by the failed attempt to prove the entailment in
\Cref{eq:lseg-ind},
fold/unfold proofs
only capture the fixpoint nature of interpretations,
and consider both finite and infinite heaps.
We refer to SLID with this relaxed semantics
as \emph{Weak Separation Logic} (\wsl{}),
formally defined in \Cref{sec:wsl}.
When theories are involved,
\wsl{} also relaxes the standard interpretation of the theory
to any interpretation satisfying its first-order axiomatization.
As a result, the class of models defining the semantics of \wsl{}
is a strict super-set of the class defining SLID.
As we show in \Cref{sec:unfolding},
\wsl{} is in fact
the logic captured by fold/unfold.
Namely, the fold/unfold proof mechanism is both sound and complete
for the theory-free, quantifier-free fragment of \wsl{}.

The choice of the weak semantics of \wsl{} is not arbitrary:
as discussed in \Cref{sec:intro},
the relaxations that culminate in \wsl{}
correspond exactly to the sources of incompleteness
of SLID.
Surprisingly,
these very same relaxations
lead to
a precise logical characterization of
\neta{cut: the completeness of}
fold/unfold proof mechanisms.

The theory-free, quantifier-free fragment of \wsl{}
is complete despite the remaining implicit second-order quantification,
which is generally a source of incompleteness.
The panacea lies in the restrictions of the fragment,
which
in conjunction with the uniqueness of heaplets
of inductive definitions,
guarantee that only finitely many partitions of heaplets
must be considered when checking entailments.
We use the same idea to generalize the completeness result
to the larger Effectively Determined-Heap (\eprsl{})
fragment of \wsl{},
where this property enables a reduction to \fol{}
and paves the way to completeness.
As we will see in the sequel,
the reduction to \fol{} is not just
a means
for proving completeness of the \eprsl{} fragment of \wsl{},
but also a cornerstone
in our approach for investigating
proof failures of fold/unfold mechanisms.

\subsection{Rogue Models}
Continuing with our study of fold/unfold proof mechanisms,
we realize that
since \wsl{} provides a logical characterization of fold/unfold reasoning,
fold/unfold proofs are made against the models of \wsl{}.
Circling back to the entailment of
\Cref{eq:lseg-ind},
we understand
that the failure of fold/unfold proofs
is due to the gap between the class of models
considered by \wsl{} and the intended class of models of SLID.
This semantic gap gives rise to entailment-refuting
\emph{rogue counter-models},
that demonstrate the root cause of failure.

A depiction of one such counter-model is given in \Cref{fig:lseg-rogue-model}.
The heap of this rogue model consists of infinitely many locations,
where $a$ is the head of an infinite sequence of locations,
each connected to the next by the points-to relation.
The sequence never reaches $c$,
but nonetheless all locations in it form a list segment with $c$,
in particular $\lseg(a,c)$ is true.
To simplify the diagram,
these are the only $\lseg$ lines depicted,
but $\lseg(x,y)$ is also true for every pair of locations
in the segment between $a$ and $c$,
where $x$ precedes $y$ in the sequence.
Other than that,
the only non-reflexive segment is between $c$ and $b$.
We can see that this is indeed a valid fixpoint
interpretation of
$\lseg$,
although not a least fixpoint (nor a greatest fixpoint).
For example, $\lseg(a,c)$ is justified by
the list segment between the successor of $a$ and $c$,
which in turn is justified by the list segment between
the successor of the successor of $a$ and $c$,
and so forth.
Since in addition $c$ points to $b$,
but $\lseg(a,b)$ is false,
it follows that the antecedent of entailment in \Cref{eq:lseg-ind} holds,
but the consequent does not.
This a rogue counter-model,
since it is neither finite nor interprets $\lseg$ as a least fixpoint.

\begin{figure}
	\centering \begin{subfigure}[t]{.485\textwidth}\centering \scalebox{0.85}{
			\begin{tikzpicture}[|->,
					>=stealth',
					shorten >=2pt,
					shorten <=2pt,
					every edge/.append style={
						line width=0.8pt,
					},
					auto,
					x=0.65cm,
					y=0.66cm,
				]
				
				\node[round] (a) at (1, 1) { $a$ };
				\node (a3) at (1, 4.5) {};
								
				\node[round] (anext) at (3, 1) {};
				\node (anext2) at (3,3.5) {};
				
				\node[round,draw=none]
				(dots) at (5, 1)
				{ $\dots$ };
				
				\node (dots1) at (5, 2.5) {};
				
				\node[round] (c) at (7, 1) {$c$};
				\node (c1) at (7, 2.5) {};
				\node (c2) at (7, 3.5) {};
				\node (c3) at (7, 4.5) {};
				
				\node[round] (b) at (9, 1) { $b$ };
				
				\node[round] (nil) at (11,1) {$\nil$};
				
				\path
					(a) edge (anext)
					(anext) edge (dots)
					(c) edge (b)
					(b) edge (nil)
				;
									
				\draw[-,line width=0.8pt]
					(dots.north) -- (dots1.center)
						-- node[above right, pos=0] {$\lseg$} (c1.center)
						-- (c.north)
						
					(anext.north) -- (anext2.center)
						-- node[above right, pos=0] {$\lseg$} (c2.center)
						-- (c.north)
						
					(a.north) -- (a3.center)
						-- node[above right, pos=0] {$\lseg$} (c3.center)
						-- ([yshift=-2pt]c.north)
				;
			\end{tikzpicture}
		}
		\caption{
			The location $a$ forms a list segment with $c$,
			justified by the infinitely many list segments that end in $c$.
			But $a$ does not form a list segment with $b$
			despite the fact that $c$ points to $b$,
			since none of the locations reachable from $a$
			(through the points-to relation)
			forms a list segment with $b$.
			One can easily verify that this is
			a fixpoint interpretation of
			\Cref{eq:lseg-sys}.
		}
		\label{fig:lseg-rogue-model}
	\end{subfigure}\hfill
	\begin{subfigure}[t]{.485\textwidth}\centering \scalebox{0.85}{
			\begin{tikzpicture}[|->,
				>=stealth',
				shorten >=2pt,
				shorten <=2pt,
				every edge/.append style={line width=0.8pt},
				auto,
				x=0.65cm,
				y=0.66cm,
				]
				
				\node[round] (a) at (1, 1) { $a$ };
				\node (a2) at (1, 3.5) {};
								
				\node[double round] (anext) at (3, 1) {};
				\node (anext1) at (3,2.5) {};
				
				\node[round] (c) at (6,1) {$c$};
				\node(c1) at (6, 2.5) {};
				\node (c2) at (6,3.5) {};

				\node[round] (b) at (8, 1) { $b$ };
				\node[round] (nil) at (10,1) {$\nil$};

				\path
				(a) edge node {$0$} (anext)
				(c) edge (b)
				(b) edge (nil)
				;
				
				\draw[|->,line width=0.8pt]
					(anext.east)
						-- ([xshift=15pt]anext.east)
						-- ([xshift=15pt,yshift=-15pt]anext.east |- anext.south)
						-- node[above left, pos=0, xshift=2pt] {$i+1$}
							([yshift=-15pt]anext.south)
						-- (anext.south)
				;
				
				\draw[-,line width=0.8pt]
					(a.north) -- (a2.center)
						-- node[above right, pos=0] {$\lseg$} (c2.center)
						-- (c.north)
						
					(c.north)
						-- ([yshift=1pt]c1.center)
						-- ([xshift=-1pt,yshift=1pt]anext1.center)
						--
							node[above right, pos=0] {$\lseg$}
							([xshift=-1pt,yshift=1pt]anext.north)
						
					(c.north)
						-- ([yshift=-1pt]c1.center)
						-- ([xshift=1pt,yshift=-1pt]anext1.center)
						-- ([xshift=1pt,yshift=-1pt]anext.north)
						
				;
			\end{tikzpicture}
		}
		\caption{
			A symbolic structure representing
			the same rogue model in \Cref{fig:lseg-rogue-model}.
			The infinitely many heap locations
			between $a$ and $c$
			are represented succinctly
			by the double-circled node in the middle,
			where each location corresponds to an
			index $i \geq 0$.
			The points-to relation connects
			$i$ to $i + 1$,
			and connects $a$ to index $0$.
}
		\label{fig:lseg-symbolic-model}
	\end{subfigure}
	\caption{A rogue counter-model to the entailment of
		\Cref{eq:lseg-ind},
		depicted both explicitly (on the left)
		and as a symbolic structure (on the right).
		Locations in the heap are shown as circles,
		$\mapsto$ arrows represent the points-to relation.
		The $\lseg$ lines show the relevant list segments ending with $c$,
		others omitted to avoid clutter.
Formally, the points-to relation is defined as
		$\braces{a \pointsi 0, c \pointsi b, b \pointsi \nil}
		\cup \midbraces{i \pointsi i + 1}{i \geq 0}$,
		and the list segment relation is defined as
		$
		\braces{
			\tuple{a,a}, \tuple{a,c},
			\tuple{c,c}, \tuple{c,b}, \tuple{c,\nil}
			\tuple{b,b}, \tuple{b,\nil}
		}
		\cup
		\midbraces{\tuple{a,i}}{i \geq 0}
		\cup
		\midbraces{\tuple{i,c}}{i \geq 0}
		\cup
		\midbraces{\tuple{i,j}}{0 \leq i \leq j}
		$.
	}
	\label{fig:lseg-both-models}
\end{figure}  
Rogue counter-models serve as a proof
of unprovability of entailments using fold/unfold,
and hint that reasoning beyond fold/unfold is necessary.
Here, the existence of an infinite list segment between $a$ and $c$
hints at a need for inductive reasoning
over the definition of $\lseg$.
Indeed, the entailment of \Cref{eq:lseg-ind}
is inductively provable,
which means that it is provable in the base case
where $a=c$,
and assuming the entailment for some list segment
$\lseg(u,c)$ such that $a \pointsi u$,
it is also provable for the list segment
$\lseg(a,c)$,
constructed
according to the inductive case of the definition.
Both the base case
and the inductive step
can be proved by fold/unfold proof mechanisms,
meaning that the only missing piece to the proof
was a single application of the induction principle,
as indicated by the rogue model.

In order for us to leverage rogue models
for understanding and fixing proof failures,
we need a way to represent and find them.
This task is challenging,
since, as seen in \Cref{fig:lseg-rogue-model},
they may be infinite.
In fact, for a large class of so-called ``heap-reducing''
inductive definitions
(which includes $\lseg$),
rogue models must be infinite,
as we prove in \Cref{sec:rogue-models}.
Fortunately, we have two helpful tools at our disposal.
The first is the \fol{} encoding of \wsl{},
which allows us to reduce the problem
of finding rogue counter-models
of SLID entailments
to finding satisfying first-order structures
for the \fo{}-encoded entailment.
The second is the recently introduced formalism of
symbolic structures~\cite{infinite-needle},
which are finite representations of possibly infinite first-order structures.
Symbolic structures enjoy decidable model-checking
and are amenable to automatic
(though unsurprisingly incomplete)
model-finding,
which, through the \fo{} encoding,
we utilize to find rogue models.

The rogue model of \Cref{fig:lseg-rogue-model}
can be represented by the symbolic structure shown in
\Cref{fig:lseg-symbolic-model}
(making the same omissions of $\lseg$ lines).
The symbolic structure compacts
the infinitely many heap locations
in the segment between $a$ and $c$
into a single symbolic \emph{summary node},
which represents each heap location with an integer $i \geq 0$,
called an index.
The points-to relation of each location $i$
represented by the summary node
is given by
$i + 1$,
meaning it points to another location
of that same summary node
that has the index $i + 1$.
The points-to relation relates $a$ to index $0$
within the summary node.
The $\lseg$ predicate treats the summary node as a whole,
interpreting all locations within it as forming a list segment with $c$.

Summary nodes give us the ability
to pinpoint the core ``rogueness'' of the model,
and hint more directly at how induction should be used.
Though in the case of \Cref{eq:lseg-ind} the entailment itself was inductive,
in other cases, induction is an intermediate step,
when an inductive lemma is used as part of a larger proof.
For example, consider the following slight variation
of \Cref{eq:lseg-ind},
where we have the added assumption that $a \neq b$:
{\smaller\begin{equation}
	\label{eq:lseg-ind-lemma}
	a \neq b * \lseg(a,c) * c \pointsi b * b \pointsi \nil \entails \lseg(a,b) * b \pointsi \nil
\end{equation}}
The entailment of \Cref{eq:lseg-ind-lemma}
is not inductive by itself,
but follows
(by fold/unfold reasoning)
from the inductively-provable entailment of \Cref{eq:lseg-ind}.
The rogue model of \Cref{fig:lseg-both-models}
is also a counter-model for \Cref{eq:lseg-ind-lemma},
and helps identify \Cref{eq:lseg-ind}
as a key lemma for completing the proof.
Firstly, it guides the user away from unhelpful lemmas and towards useful ones:
any entailment that already holds in the counter-model
would not help us eliminate it,
and thus we can focus on entailments not satisfied by the rogue model.
I.e., entailments where the antecedent holds in the model
but the consequent does not.
In the case of \Cref{fig:lseg-both-models},
one can see that the entailment
$\lseg(a,c) * c \pointsi b * b \pointsi \nil \entails \lseg(a,b) * b \pointsi \nil$ 
of \Cref{eq:lseg-ind}
is one such possibility, since all the atoms in the antecedent are satisfied
but at least one of the atoms in the consequent is not.
Secondly, the rogue model also hints where induction should be used.
We can see that though $\lseg(a,c)$ is true in the rogue model,
the nodes of $a$ and $c$ are not connected in the symbolic structure
according to the points-to relation.
This means that $\lseg(a,c)$ cannot be justified in a 
least-fixpoint interpretation,
and thus induction on this term could prove useful.
These two sources of guidance can be used somewhat systematically:
mapping out formulas that are satisfied and not satisfied by the model
to construct lemmas,
and looking for inductive-predicate atoms that ``close over'' infinite parts
of the model (that manifest as disconnected in the symbolic representation)
to inform induction.

For a more elaborate example consider
a system of inductive definitions that uses
the background theories of integers and sets to model (strictly) sorted lists
and their keys,
where we have two inductive predicates $\slist(x)$, for sorted list,
and $\keys(x, Y)$, for the set of keys in elements reachable from $x$:
{\smaller\[
\tightarray
\begin{array}{rcl}
	\slist(x) &\coloneq &
		x = \nil \lor
		\parens{\exists u. x \pointsi u * \slist(u) * \key(x) < \key(u)}
	\\
	\keys(x, Y) &\coloneq& (x = \nil \land Y = \emptyset) \lor
		\parens{\exists u, Z. x \pointsi u * \keys(u, Z) * Y = Z \cup \braces{\key(x)}}
	,
\end{array}
\]}
and the following entailment:
{\smaller\[
a \pointsi b * \key(a) < \key(b)
	* \parens{\slist(b) \land \keys(b, K)}  \vdash \slist(a) * \key(a) \notin K
.\]}
The entailment is valid in SLID but not in WSL 
since WSL allows (rogue) models 
where $b$ is the head of an infinite sorted list
and $\key(a)$ is in the set of keys reachable from it
even though $\key(a) < \key(b)$.
For example, if we identify each element in the heap with its key,
one such rogue model can be described with the following points-to relation:
$a = 0 \pointsi b = 1 \pointsi 2 \pointsi \dots$;
where $\slist(i)$ is true for all $i$;
and $\keys(i, K)$ is true iff $K = \braces{0, i, i+1, i+2, \dots}$. 
Looking at this model,
it is clear that for any $\keys(i, K)$ where $i > 0$,
$K$ spuriously contains the element 0.
This interpretation of $\slist$ and $\keys$ is a valid fixpoint but not a least fixpoint, 
where a key is only included in the set of reachable keys if a
corresponding element is in the list. Since the keys keep
increasing in a sorted list, this ensures that in a
least-fixpoint interpretation we will never encounter a key
smaller than the key of the head of the list. In particular,
in any least-fixpoint interpretation where $\key(a) <
\key(b)$, $\slist(b)$ and $\keys(b, K)$ hold, we expect
$\key(a) < k$ for any $k \in K$. This property, which is
violated in the rogue model,  
can be written as an entailment that is provable by
induction on $\slist(b)$ and implies the desired property.
The rogue model not only helps identify this entailment as
an auxiliary lemma, but also hints at induction on
$\slist(b)$ for proving it since $b$ is disconnected from
$\nil$ (the base case of $\slist$). Representing this kind
of rogue model, where multiple background theories are
involved, requires careful modeling, but is possible in
theory.

The use of symbolic structures to represent rogue models
is not merely a theoretical curiosity.
As we discuss in \Cref{sec:eval},
we developed a prototype tool that implements
the \fo{} encoding of \wsl{}
with the background theory of linear integer arithmetic,
and uses \fest{}~\cite{infinite-needle}
to find symbolic rogue counter-models.
In parallel, the tool uses \zzz{}~\cite{Z3}
to check the validity of the \fol{} formulas resulting from the encoding, which implies validity in \wsl{} and therefore in SLID.
We evaluated
the effectiveness of this approach
over a benchmark suite consisting of over 700 examples
of entailments with inductive definitions,
taken from the \sls{} tool for
inductive lemma synthesis in SLID~\cite{sls}.
Our tool is able to find infinite rogue models
for up to 92\% of the examples
that require reasoning beyond fold/unfold,
and provide helpful feedback for understanding proof failures.

\section{Preliminaries}
\label{sec:prelim}

\paragraph{First-Order Logic}
	We define many-sorted first-order logic (\fol{}) 
	with equality in the usual way.
	A first-order vocabulary consists of sorted
	constant, function and relation symbols.
	We denote the set of logical variables 
	$x, y, x', x_1, \dots$
	by $\Vars$,
	and assume each variable has an associated sort,
	omitted when clear from context.
	Terms are constructed from constants, variables
	and well-sorted function applications.
	Formulas are built up from equalities and relation applications
	using logical connectives $\neg, \lor, \land, \to$
	and quantifiers $\forall, \exists$.
	
\smallpara{Background theories and \lia{}}
	In literature, background theories are usually defined either 
	as theories of a standard model or by a set of axioms. 
	We consider these two faces together,
	and accordingly define a theory as a triple
	$\bg = \parens{ \Sigma^\bg, M^\bg, \Gamma^\bg }$
	where $\Sigma^\bg$ is a first-order vocabulary,
	$M^\bg$ is a first-order structure for
	$\Sigma^\bg$,
	and $\Gamma^\bg$ is a recursive set of first-order
	sentences over $\Sigma^\bg$
	that is sound for $M^\bg$
	(i.e., $M^\bg \models \Gamma^\bg$).
	We call
	$\Sigma^\bg$ the theory vocabulary,
	$M^\bg$ the
	\emph{standard model} of \bg{},
	and $\Gamma^\bg$ 
	the first-order axiomatization of \bg{}.
In particular we consider the theory of Linear Integer Arithmetic
$\intTheory = \parens{ \intVocab, \intModel, \intAxioms }$,
where $\intVocab = \braces{ 0, 1, +, < }$
over sorts $\braces{\intSort}$,
$\intModel$ is the standard model of the integers,
and $\intAxioms$ is the complete first-order axiomatization
of \lia{}.
We denote the set of quantifier-free \lia{} formulas by $\qfLia$.

\section{A Weak Separation Logic}
\label{sec:wsl}

In this section we formally define Weak Separation Logic (\wsl{}),
a relaxed version of
separation logic with inductive definitions
and theories,
that tackles its sources of incompleteness.
For brevity, from now on we use \seplog{}
also when referring to SLID,
and let context clarify whether inductive definitions
are involved.
We establish completeness for a fragment of \wsl{}
by reduction to first-order logic.
In \Cref{sec:unfolding}
we further show that \wsl{}
captures the actual semantics that fold/unfold
proof mechanisms for \seplog{}
are sound and complete for.
The class of satisfying WSL models is a strict superset
of the class of satisfying SL models,
giving rise to \emph{rogue, nonstandard models},
which we discuss in \Cref{sec:rogue-models}.

\subsection{Separation Logic with Inductive Definitions
	and Theories}

We start by giving a model-oriented definition of \seplog{}.
This allows us to analyze both \seplog{} and \wsl{}
through their classes of models,
thereby getting better understanding
of the semantic gap between them.
\neta{improve}
In literature, the semantics of \seplog{}
is usually\neta{conventionally?} given in relation
to a store $s$ and a heap $h$
that together describe a program state.
Our model-oriented description uses different objects,
namely heap structures $M$, assignments $v$ and heaplets $\eta$,
to describe a state, but the two representations
and corresponding semantics are equivalent.

From this point on,
and throughout the paper,
we fix a background theory
$\bg = \parens{\Sigma^\bg, M^\bg, \Gamma^\bg}$
over a set of sorts $S$.
\neta{$S$ or $S^\bg$? We use $S$
	but we can migrate to $S^\bg$}
We start with the basic vocabulary and syntax of \seplog{} formulas.

\begin{definition}[Vocabulary of Separation Logic]
	A vocabulary of SL
	over background theory \bg{}
	 is a set
	$\Sigma =
	\Sigma^\fg \uplus \Sigma^\bg$,
	where
	$\Sigma^\bg$ is the background vocabulary
	over sorts
	$S$, and
	$\Sigma^\fg$
	is a \emph{relational} first-order vocabulary over
	$S \uplus \braces{\fgSort}$
	(i.e., containing only constant and relation symbols),
	required to contain a constant symbol $\nil$ of sort $\fgSort$.
	We call $\fgSort$ the \emph{location sort},
	and the predicates in $\Sigma^\fg$ the
	\emph{inductive predicates}.
	By convention, we use $P, P', P_1, \dots$ for inductive predicates
	and $Q, Q', Q_1, \dots$ for theory predicates.
\end{definition}

The syntax of \seplog{} formulas contains an atomic ``points-to'' relation,
which relates locations to their contents.
In order to simplify the presentation,
and in particular remove uninteresting
details from the \fol{} reduction,
we assume a uniform, global, \emph{record shape},
a tuple of sorts $\tuple{\sort_1, \dots, \sort_n}$
that every location adheres to.
This assumption is purely for presentation reasons,
and in fact all results hold just the same
for a set of record shapes.

\begin{definition}[Syntax of  SL formulas]
	\label{def:sl-syntax}
	Given a
	vocabulary
	$\Sigma = \Sigma^\fg \uplus \Sigma^\bg$
	and a record shape
	$\tuple{\sort_1, \dots, \sort_n}$
	over $S \uplus \braces\fgSort$,
	terms and formulas are defined recursively,
	by the following grammar:
\[
	\tightarray
	\begin{array}{rcl}
		t &\Coloneq &
			c \in \Sigma
			\spacedMid x \in \Vars
			\spacedMid f(t_1, \dots, t_k)
		\\
		\varphi &\Coloneq&
			t_1 = t_2
			\spacedMid t_1 \neq t_2
			\spacedMid Q(t_1, \dots, t_k)
			\spacedMid \neg Q(t_1, \dots, t_k)
			\spacedMid \emp
			\spacedMid t \points {t_1, \dots, t_n}
			\spacedMid P(t_1, \dots, t_k)
		\\
			&\mid& \varphi_1 \lor \varphi_2
			\spacedMid \varphi_1 \land \varphi_2
			\spacedMid \varphi_1 * \varphi_2
			\spacedMid \exists x. \varphi
			\spacedMid \forall x. \varphi
	\end{array}
	\]
	We
use the names
	\emph{theory-free} for formulas where no theory terms\sharon{or pedicates? or is it implied?} appear,
	\emph{quantifier-free} for formulas with no quantifiers,
	and \emph{conjunctive} for formulas without disjunctions
	(but with both regular and separating conjunctions).
\end{definition}

Though under standard SL semantics
we are interested only in structures
that interpret the inductive predicates
as a least-fixed-point solution
of a system of inductive definitions,
in order to define this formally,
we first define a satisfaction relation for
triples of structure $M$,
assignment $v$
and heaplet $\eta$,
then restrict our attention to valid \lfp{} structures.
Such a triple $M, v, \eta$
is simply a different way of representing
the typical store and heap of \seplog{} semantics.
The store is split between interpretations of constants $c^M$
and assignment to variables $v(x)$,
the heap is split between a global heap mapping
$h^M$ that represents the heap's points-to relation
and the heaplet $\eta$
that describes the heap's domain,
i.e., the set of allocated locations.
The heap can therefore be reconstructed from $h^M$ projected onto $\eta$.
The satisfaction relation defined in \Cref{def:sl-sat}
then coincides with satisfaction by a store-heap pair.

\begin{definition}[Heap structures]
	Given a  vocabulary
	$\Sigma = \Sigma^\fg \uplus \Sigma^\bg$
	and a record shape
	$\tuple{\sort_1, \dots, \sort_n}$
	over $S \uplus \braces\fgSort$,
	a \emph{heap structure} for
	$\Sigma, \tuple{\sort_1, \dots, \sort_n}$
	is a triple
	$M = \parens{\Dom, h, \Interp}$,
	where $\Dom$ is a mapping of sorts to disjoint domains,
	of which $\Dom(\fgSort)$,
	called a \emph{location domain},
	contains a designated element
	$\Null \in \Dom(\fgSort)$;
	$h$ is a \emph{heap}, a mapping
	of non-null locations to records
	of domain elements,
	$h \colon \Dom(\fgSort)_+
	\to \Dom(\sort_1) \tdots \Dom(\sort_n)$,
	where $\Dom(\fgSort)_+ \defeq
	\Dom(\fgSort) \setminus \braces\Null$;
	and $\Interp$ is an \emph{interpretation} of vocabulary symbols,
	defined in the usual way,
	except that for an inductive predicate symbol
	$P \colon \sort_1 \tdots \sort_k$,
	the interpretation is defined as
	$\Interp(P) \subseteq \Dom_P \defeq
		\Dom(\sort_1) \tdots \Dom(\sort_k)
		\times \Power\parens{\Dom(\fgSort)}
		$.
	Additionally, we require that
	$\Interp(\nil) = \Null$.
	For any symbol $s$, we denote
	$s^M = \Interp(s)$.
		
	A structure $M$ is called \emph{heap-finite} if
	its location domain $\Dom(\fgSort)$ is finite.
	We define a heaplet $\eta$ as a set of non-null locations, i.e.,
	$\eta \subseteq \Dom(\fgSort)_+$,
	and an assignment
	as a well-sorted mapping $v \colon \Vars \to \Dom$
	and we denote by $\bar v$ the lifting of $v$ to any term
	$t$ in the usual way.

\end{definition}

\begin{definition}[Satisfaction of SL formulas]
	\label{def:sl-sat}
	Given a heap structure $M$, an assignment $v$
	and a heaplet $\eta$,
	we recursively define satisfaction of a SL formula
	in \Cref{fig:slrd-formula-sat}.
\end{definition}

\begin{figure}
	\[
	\begin{array}{rcllrrl}
		M, v, \eta
		& \models &
		t_1 \bowtie t_2
		& \text{if} &
		\bar v(t_1) \bowtie \bar v(t_2)
		& \text{and} & \eta = \emptyset
		\\
		M, v, \eta
		& \models &
		Q(t_1, \dots, t_k)
		& \text{if} &
		\tuple{\bar v(t_1), \dots, \bar v(t_k)} \in Q^M
		& \text{and} &
		\eta = \emptyset
		\\
		M, v, \eta
		& \models &
		\neg Q(t_1, \dots, t_k)
		& \text{if} &
		\tuple{\bar v(t_1), \dots, \bar v(t_k)} \notin Q^M
		& \text{and} &
		\eta = \emptyset
		\\
		M, v, \eta
		& \models &
		\emp
		& \text{if} &
		&&
		\eta = \emptyset
		\\
		M, v, \eta
		& \models &
		P(t_1, \dots, t_k)
		& \text{if} &
		\tuple{\bar v(t_1), \dots, \bar v(t_k), \eta} \in P^M
		\\
		M, v, \eta
		& \models &
		t \points{t_1, \dots , t_n}
		& \text{if} &
		\bar v(t) \neq \Null,
		h^M \parens{ \bar v (t) } =
		\tuple{ \bar v (t_1), \dots, \bar v (t_n) }
		& \text{and} &
		\eta = \braces{ \bar v(t) }
		\\
		
		M, v, \eta
		& \models &
		\varphi_1 \lor \varphi_2
		& \text{if} &
		M, v, \eta \models \varphi_1
		\text{ or }
		M, v, \eta \models \varphi_2
		\\

		M, v, \eta
		& \models &
		\varphi_1 \land \varphi_2
		& \text{if} &
		M, v, \eta \models \varphi_1
		\text{ and }
		M, v, \eta \models \varphi_2
		\\

		M, v, \eta
		& \models &
		\varphi_1 * \varphi_2
		& \text{if} &
		\exists \eta_1, \eta_2.\;
		M, v, \eta_1 \models \varphi_1,
		M, v, \eta_2 \models \varphi_2
		& \text{and} &
		\eta = \eta_1 \uplus \eta_2
		\\
		M, v, \eta
		& \models &
		\mathcal{Q} x \colon \sort. \varphi
		& \text{if}
		& \Quantifier d \in \Dom^M(\sort).
		M, v\SingSubstitute{x}{d}, \eta \models \varphi
	\end{array}
	\]
	\Description{Recursive definition of satisfaction in Separation Logic}

	\caption{Recursive definition of satisfaction in Separation Logic.
		The $\bowtie$ symbol stands for either $=$ or $\neq$,
		the $\Quantifier$ symbol stands for either $\forall$ or $\exists$,
		and $\eta = \eta_1 \uplus \eta_2$ is shorthand for
		$\eta = \eta_1 \cup \eta_2 \land \eta_1 \cap \eta_2 = \emptyset$.
	}
	\label{fig:slrd-formula-sat}
\end{figure}

Using the definitions of heap structures
and formula satisfaction,
we now move to formally
define systems of inductive definitions
and their semantics.

\begin{definition}[System of inductive definitions]
	\label{def:sid-syntax}
	Given a vocabulary $\Sigma$,
	a \emph{system of inductive definitions} (SID) $\sys$
	is a mapping of each inductive predicate
	$P \colon \sort_1 \tdots \sort_k$,
	to a formula
	$\sys(P)$ with free variables $\Vec x = x_1, \dots, x_k$,
of the form
	$\sys(P)(\Vec x) = \exists \Vec{y}. \biglor_j \rho^P_j(\Vec x, \Vec y)$,
	where
$\rho^P_j$ is a quantifier-free, conjunctive \seplog{} formula.

	Note that in literature,
	the inductive cases $\rho^P_j$
	are commonly defined as containing the existential quantifier.
	This is equivalent (under variable renaming),
	but we prefer this form for the simplicity of presentation.
\end{definition}

\begin{notation}
	Given a structure $M$ over a vocabulary
	with inductive predicates
	$\Vec P = \tuple{P_1, \dots, P_m}$,
	and a tuple of predicate interpretations
	$\Vec I = \tuple{I_1, \dots, I_m}$,
	where $I_i \subseteq \Dom_{P_i}$,
	we denote the reinterpretation model
	$M' = M \SingSubstitute{\Vec{P}}{\Vec{I}}$
	where $P_i^{M'} = I_i$.
\end{notation}

\begin{definition}[Transformer
	for system of inductive definitions]
	\label{def:sid-sat}
	Given a structure $M$ over a vocabulary
	with inductive predicates
	$P_1, \dots, P_m$,
	and a system of inductive definitions $\sys$,
	for each relation
	$P \colon \sort_1 \tdots \sort_k$,
	we define an operator
	$\sys^M_P \colon \Dom_{P_1} \times \dots \times \Dom_{P_n} \to \Dom_P$:
	\[
	\sys^M_P \parens{\Vec{I}} \defeq \midbraces{
		\tuple{ d_1, \dots, d_k, \eta }
	}{
		\begin{array}{c}
			d_i  \in \Dom^M(\sort_i), \\
			\eta \subseteq \Dom^M(\fgSort), \\
			M\SingSubstitute{\Vec P}{\Vec{I}},
			\SingSubstitute{\Vec{x}}{\Vec{d}},
			\eta
			\models
			\sys(R)
		\end{array}
	};
	\]
	and we define a transformer for the entire system:
	$\sys^M \colon \Dom_{P_1} \times \dots \times \Dom_{P_n} \to
	\Dom_{P_1} \times \dots \times \Dom_{P_n}$:
	\[
	\sys^M \parens{\Vec I} \defeq \tuple{
		\sys^M_{P_1}\parens{\Vec I},
		\dots,
		\sys^M_{P_n}\parens{\Vec I}
	}.
	\]
	
	Note that $\sys^M(\Vec I)$ is monotone with respect to
	set inclusion
	(since all inductive predicates appear positively),
	and therefore has a least fixpoint~\cite{tarski}.
	We say that $M$ is a \fp{} structure for $\sys$ when
	$\Vec{P}^M = \sys^M\parens{\Vec{P}^M}$
	and that $M$ is a \lfp{} structure for $\sys$ when
	$\Vec{P}^M$ is the least set such that
	$\Vec{P}^M = \sys^M\parens{\Vec{P}^M}$.
	We denote the classes of all \fp{}, \lfp{} structures for $\sys$
	as $\Models_\fp(\sys), \Models_\lfp(\sys)$ respectively.
\end{definition}

Given a heap structure $M$ and
$\tuple{d_1,\ldots,d_k, \eta} \in P^M$,
we refer to $\eta$ as a supporting heaplet of
$\tuple{d_1,\ldots,d_k}$.
In general, a \lfp{} or \fp{} structure allows the same tuple of elements
to have multiple supporting heaplets for the same inductive predicate.
However, as explained in \Cref{sec:intro},
this is both an obstacle for overcoming incompleteness resulting
from second-order quantification over heaplets,
and defies the design goal of separation logic,
where formulas are meant to characterize heap locations.
Indeed, typical SIDs ensure that in their \lfp{} structures,
every tuple of elements that satisfies an inductive predicate
has a unique heaplet (often called footprint) that supports it.
Next we formalize this property of structures,
and restrict our attention to systems that guarantee it.

\begin{definition}[Determined-heap structure]
	\label{def:detheap-struct}
	A structure $M$ has a \emph{determined heap}
	if for any inductive predicate
	$P \colon \sort_1 \tdots \sort_k$
	and for any tuple of elements $d_1, \dots, d_k$
	there exists at most one heaplet
	$\eta \subseteq \Dom(\fgSort)$
	s.t.
	$\tuple{d_1, \dots, d_k, \eta} \in P^M$.
	For a determined-heap structure we may denote
	the interpretation of  inductive predicate as a partial function
	$P^M \colon \Dom(\sort_1) \tdots \Dom(\sort_k) \partto \Power(\Dom(\fgSort))$,
	mapping tuples of elements to their supporting heaplet.
\end{definition}

We round this section off by
summarizing the syntax and semantics of \seplog{},
and defining two kinds of entailment in \seplog{}:
plain, and modulo a system of inductive definitions.

\begin{definition}[Separation Logic]
	\label{def:sl}
	We define Separation Logic (\seplog{})
	as a logic with the syntax of formulas
	defined in \Cref{def:sl-syntax},
	the syntax of inductive definitions
	defined in \Cref{def:sid-syntax},
	and semantics given as the satisfaction relation
	\Cref{def:sl-sat},
	over the class $\Models_\seplog$
	of \emph{heap-finite} determined-heap structures
	(\Cref{def:detheap-struct}),
	that are extensions of the standard model of the background theory
	$M^\bg$.
	
	For a finite set of \seplog{} formulas $\Gamma$
	and a \seplog{} formula $\psi$,
	we say that the \seplog{} entailment problem
	for $\Gamma$ and $\psi$,
	denoted $\Gamma \entails_\seplog \psi$,
	is valid
	when for every model
	$M \in \Models_\seplog$,
	assignment $v$ and heaplet $\eta$,
	if $M, v, \eta \models \Gamma$ 
	then $M, v, \eta \models \psi$.
Further, we say that
	the \seplog{} entailment problem for $\Gamma$ and $\psi$
	modulo SID $\sys$,
	denoted	$\Gamma \sysentails_\seplog \psi$,
	is valid
	when for every model
	$M \in \Models_\seplog \cap \Models_\lfp(\sys)$,
	assignment $v$ and heaplet $\eta$,
	if $M, v, \eta \models \Gamma$ 
	then $M, v, \eta \models \psi$.
\end{definition}

\subsection{From Separation Logic to Weak Separation Logic}
As discussed in \Cref{sec:overview}, \seplog{} contains
multiple sources of incompleteness,
which makes it difficult to discern the root cause of a proof failure,
be it a deficient proof or an unprovable entailment.
In this section we define Weak Separation Logic (\wsl{}),
that relaxes the semantics of \seplog{}
and allows us to understand proof failures of \seplog{}
as stemming from the semantic gap between \seplog{} and \wsl{}.

\wsl{} tackles three of the four sources of incompleteness in \seplog{}
with three relaxations that extend the class of structures considered:
\begin{enumerate*}[(i)]
	\item
	structures are not necessarily extensions of  the standard model of the theory,
	but only required to satisfy its axiomatization;
	\item
	the location domain may be infinite
	(not just finite); and
	\item
	for entailments modulo systems of inductive definitions,
	we use \fp{} structures
	(not just \lfp{} structures).
\end{enumerate*}
Formally:

\begin{definition}
	We define Weak Separation Logic (\wsl{})
	with the same syntax of formulas and inductive definitions
	as in \seplog{},
	and semantics given as the same satisfaction relation
	as in \seplog{},
	but over the class $\Models_\wsl$
	of \emph{heap-finite or infinite} determined-heap structures
	that \emph{satisfy $\Gamma^\bg$}.
	
	For a finite set of \seplog{} formulas $\Gamma$
	and \seplog{} formula $\psi$,
	we say that the \wsl{} entailment problem for $\Gamma$ and $\psi$,
	denoted $\Gamma \entails_\wsl \psi$,
	is valid
	when for every model
	$M \in \Models_\wsl$,
	assignment $v$ and heaplet $\eta$,
	if $M, v, \eta \models \Gamma$ 
	then $M, v, \eta \models \psi$.
Further, we say that
	the \wsl{} entailment problem for $\Gamma$ and $\psi$
	modulo SID $\sys$,
	denoted	$\Gamma \sysentails_\wsl \psi$,
	is valid
	when for every model
	$M \in \Models_\wsl \cap \Models_\fp(\sys)$,
	assignment $v$ and heaplet $\eta$,
	if $M, v, \eta \models \Gamma$ then $M, v, \eta \models \psi$.
\end{definition}

Recall that the remaining source of incompleteness in \seplog{}
is the implicit second-order quantification over heaplets.
\wsl{} does not tackle this source directly,
but we later consider a fragment of \wsl{},
where restricted quantification together with determined-heap
property of structures
allow us to overcome this obstacle.

We now focus our discussion
on the relationship between \seplog{} and \wsl{}.
\wsl{} is clearly a relaxation of \seplog{},
in the sense entailments that are valid in \wsl{}
are also valid in \seplog{},
whether plain or modulo some SID,
but the converse does not necessarily hold.
However, next we show that when restricting ourselves to entailments
where \seplog{} does not exhibit the sources of incompleteness,
\seplog{} and \wsl{} coincide,
which means that in these cases
the relaxations are inconsequential.

Intuitively, to eliminate incompleteness stemming from the theory,
we restrict ourselves to the sub-logics of \seplog{} and \wsl{}
without theories.
To remove the distinction between finite and infinite models
and handle second-order quantification over heaplets,
we restrict ourselves to QF formulas,
which enjoy a finite-model property.
Together these restrictions ensure that for every QF formula $\varphi$,
there exist $M \in \Models_\seplog$,
assignment $v$ and heaplet $\eta$
s.t. $M, v, \eta \models \varphi$
iff
there exist $M' \in \Models_\wsl$, $v'$ and $\eta'$
s.t. $M', v', \eta' \models \varphi$.
Finally, to eliminate incompleteness stemming from the gap between
\lfp{} and \fp{} semantics,
we consider plain entailments,
i.e., without systems of inductive definitions.
Indeed we show that under these restrictions
the semantic gap between \seplog{} and \wsl{} is eliminated.

\begin{claim}
	\label{thm:sl-wsl-equi-valid}
	Given a finite set of theory-free, quantifier-free formulas $\Gamma$
	and a theory-free, quantifier-free formula $\psi$,
	the entailment
	$\Gamma \entails_\seplog \psi$ is valid
	iff
	$\Gamma \entails_\wsl \psi$ is valid.
\end{claim}

\Proof{thm:sl-wsl-equi-valid}{
	As shown in~\cite{results-for-spatial-assertion-language},
	theory-free, quantifier-free formulas in \seplog{}
	enjoy a small-model property.
	This property is defined in relation to the satisfaction relation
	$M, v, \eta \models \varphi$,
	which is shared between \seplog{} and \wsl{}.
	As $\Gamma$ itself is finite,
	the validity of the entire entailment $\Gamma \entails_\Logic \psi$
	can be checked by considering only finite models,
	where the semantics of \seplog{} and \wsl{} coincide.
}

\Cref{thm:sl-wsl-equi-valid} implies that proof systems
for \seplog{} and \wsl{} coincide under these restrictions, and therefore the logics are ``as complete'' in this setting.
In fact, it is not difficult to see that both logics are complete under these restrictions since they enjoy a small-model property,
which means that checking the validity of theory-free, quantifier-free
\seplog{} entailments and  \wsl{} entailments is even decidable:

\begin{claim}
	\label{thm:qf-sl-decidable}
	The validity problem of theory-free, quantifier-free entailments (without SIDs)
	is decidable for both
	\seplog{} (\cite{results-for-spatial-assertion-language})
	and \wsl{}.
\end{claim}

\Proof{thm:qf-sl-decidable}{
	As in the proof of \Cref{thm:sl-wsl-equi-valid},
	we follow~\cite{results-for-spatial-assertion-language}
	and use the small-model
	property of theory-free, quantifier-free formulas.
	Checking validity (for either \seplog{} or \wsl{}) can be decidably done
	by iterating models for the antecedent and consequent up to the bound
	given in~\cite{results-for-spatial-assertion-language}.
}

The correspondence between \seplog{} and \wsl{} breaks down once theories,
systems of inductive definitions and even limited quantification
are involved,
since \seplog{} becomes incomplete with the inclusion of any of them,
but as we will prove in the next subsection, \wsl{} remains complete even when all three are included.

\subsection{Completeness of
	Effectively Determined-Heap
	WSL
	 via First-Order Encoding}
\label{sec:fo-encoding}

In this subsection we define a fragment of \wsl{},
called Effectively Determined-Heap (\eprsl{}),
where the second-order quantification over heaplets
collapses to a single determined heaplet
for every formula in the context of entailments.
We then give a first-order encoding of
\wsl{} entailments of \eprsl{} formulas
modulo systems of inductive definitions,
and show that it is \emph{model preserving}:
that is, we can capture precisely the semantics
of \wsl{} in \fol{}.
Our \fo{} encoding resembles existing encodings of \seplog{}
(e.g.,~\cite{separation-predicates}).
However, prior work does not show completeness properties
for the encoding,
and caters to restricted fragments,
where SIDs are required to have
distinct cases in the definition
of each inductive predicate.

\begin{definition}
	The fragment of
	Effectively Determined-Heap (\eprsl{})
	formulas
	consists of formulas
	of the form
	$\exists \Vec y. \biglor_i \forall \Vec u_i. \varphi_i$,
	where each $\varphi_i$ is a quantifier-free \emph{conjunctive} formula.
	We denote by \eprwsl{} the sub-logic of \wsl{},
	where formulas are restricted to the \eprsl{} fragment.
\end{definition}

As mentioned above, \wsl{}
does not directly tackle the incompleteness stemming
from the second-order quantification over heaplets.
Instead, this source of incompleteness is eliminated
by the syntactic restrictions of the \eprwsl{} fragment.
\eprwsl{} is defined in a manner similar to
the Effectively PRopositional (EPR)
fragment of \fol{},
but whereas EPR only restricts the quantifier structure,
\eprwsl{} also restricts disjunctions in formulas.\sharonnew{On the other hand, while EPR forbids $\forall^*\exists^*$ quantification, \eprwsl{} may exhibit such quantification due to the existential quantification allowed in definitions of inductive predicates.}
While EPR ensures that for any formula,
it suffices to consider finite structures
for establishing (un)satisfiability,
\eprwsl{} does not ensure that.
Instead,
it ensures that for any formula and structure,
it suffices to consider finitely many heaplets
for establishing validity of entailments.
These properties make EPR a decidable fragment of \fol{},
while
\eprwsl{} is an undecidable fragment of \wsl{} (\Cref{thm:wsl-undecidable})
that admits a complete proof system (\Cref{thm:wsl-complete}).

Note that the restrictions imposed by the \eprsl{} fragment
do not suffice to ensure completeness in \seplog{}.
In particular, \seplog{} is incomplete already
for theory-free, quantifier-free entailments modulo SIDs,
which are included in \eprsl{}.

\neta{acronym UC for universal conjunctive?
	conjunctive universal?}
	
\begin{remark}
	The \eprsl{} fragment resembles the notion of \emph{precise predicates}
	(introduced in~\cite{ohearn04}),
	but differs in two important ways.
	Firstly, \eprsl{} is a syntactic fragment of formulas,
	whereas precision is a semantic property of predicates and formulas:
	using our terminology, a formula $\varphi$ is precise if for any
	structure $M$ and assignment $v$
	there exist at most
	one heaplet $\eta$ such that $M, v, \eta \models \varphi$.
	Secondly, the two notions are incomparable.
	Indeed,
	not all precise formulas are in \eprsl{},
	and the formulas we allow in \eprsl{} need not be precise,
	since \eprsl{} ensures that there are only finitely
	many satisfying heaplets but not necessarily just one.
	For example, the formula
	$\forall x. \exists y. x = y$
	does not meet
	the syntactic requirements of \eprsl{}
	but it is precise.
	Conversely, the formula
	$\emp \lor x \points{\nil}$
	is included in \eprsl{},
	but it can be satisfied by two heaplets
	($\emptyset$ and $\braces{x}$),
	and therefore it is not precise.
\end{remark}

\paragraph{Encoding of \eprwsl{} entailments in \fol{}}
We first observe that we can reduce an entailment
$\Gamma \sysentails_\wsl \psi$
where $\Gamma \cup \braces \psi$ consists of \eprsl{} formulas,
to a finite set of entailments
of the form
$\varphi \sysentails_\wsl \psi$
where $\varphi$ is a universal conjunctive formula
and $\psi$ remains unchanged,
such that $\Gamma \sysentails_\wsl \psi$ is valid
iff all the resulting entailments are valid.
To that end, we first skolemize
$\Gamma$, eliminating the existential quantifiers,
and by using distributivity
over the conjunction $\bigland \sk (\Gamma)$,
we translate it to an equivalent
disjunction of universal conjunctive formulas.
Then,
by relying on the fact that $\biglor_i \varphi_i \sysentails_\wsl \psi$
is valid
iff $\varphi_i \sysentails_\wsl \psi$ is valid for every $i$,
we obtain a finite set of entailments of the aforementioned form.

In the sequel we show how to encode
an entailment of the form $\varphi \sysentails_\wsl \exists^* \biglor \psi_i$,
where $\varphi, \psi_i$ are universal, conjunctive formulas to \fol{}.
A key idea of the translation is to
\begin{enumerate*}[(i)]
	\item
	replace
	the points-to relation
	with unary functions $m_1, \dots, m_n$,
	each of which captures one field
	in the record a heap location points to; and
\item
	replace
each inductive predicate by a pair of \fo{} predicates,
that together represent the same mapping.
Recall, that in determined-heap structures,
every inductive predicate $P$
is interpreted as  a partial function
from tuples of elements,
that satisfy the predicate,
to the (unique) supporting heaplet.
In the \fo{} encoding we use $P_\fo$ to represent
the domain of $P$,
i.e., the set of tuples for which the predicate holds,
and $P_\eta$ represents the range of $P$,
i.e., for each tuple, the heaplet supporting its inclusion in the predicate.
Instead of mapping each tuple to its supporting heaplet,
$P_\eta$ represents this mapping as a relation between tuples
and locations in their heaplet.
\end{enumerate*}

\begin{definition}[Structure correspondence]
	Given a vocabulary
	$\Sigma = \Sigma^\fg \uplus \Sigma^\bg$
	and record shape $\tuple{\sort_1, \dots, \sort_n}$
	over sorts $S \uplus \braces\fgSort$,
	we define a FO-corresponding vocabulary
	$\FOEncoding(\Sigma) = \Sigma^\fg_\fo \uplus \Sigma^\bg$
	over $S \uplus \braces\fgSort$,
	where $\Sigma^\fg$ and $\Sigma^\fg_\fo$
	contain the same constant symbols;
	for every $i \in \brackets n$,
	$\Sigma^\fg_\fo$ contain a function symbol
	$m_i \colon \fgSort \to \sort_i$;
	and for every inductive predicate $P \in \Sigma^\fg$
	with signature $\sort_1, \dots, \sort_k$
	there exists a pair of relations $P_\fo, P_\eta \in \Sigma^\fg_\fo$
	with signatures
	$P_\fo \colon \sort_1 \tdots \sort_k,
	P_\eta \colon \sort_1 \tdots \sort_k \times \fgSort$.
		
	We define a correspondence
	between WSL structures over $\Sigma$
	and FO structures over $\FOEncoding(\Sigma)$.
	Two structures $M, M_\fo$
	over $\Sigma, \FOEncoding(\Sigma)$ (resp.)
	are said to correspond when
	\begin{enumerate}
		\item
		they have the same domain
		$\Dom^M = \Dom^{M_\fo}$;
		
		\item
		for any symbol $s \in \Sigma^\bg$,
		$s^M = s^{M_\fo}$;
		
		\item
		for any constant symbol $c \in \Sigma^\fg$,
		$c^M = c^{M_\fo}$;
		
		\item
		for any element $d \in \Dom(\fgSort)$,
		$h^M(d) = \tuple{m^{M_\fo}_1(d), \dots, m^{M_\fo}_n(d)}$;
		and
		
		\item
		for any inductive predicate
		$P \colon \sort_1 \tdots \sort_k$
		and elements $d_1 \in \Dom^M(\sort_1), \dots, d_k \in \Dom^M(\sort_k)$,
{\smaller\[
		P^M(d_1, \dots, d_k) = \left\{
		\begin{array}{ll}
			\midbraces{ d' }{ \tuple{d_1, \dots, d_k, d'} \in P^{M_\fo}_\eta }
			& \text{if } \tuple{d_1, \dots, d_k} \in P^{M_\fo}_\fo,
			\\
			\bot & \text{otherwise.}
		\end{array}
		\right.
		\]}
	\end{enumerate}
\end{definition}

Note that every \wsl{} structure $M$ over $\Sigma$
has a corresponding \fo{} structure $M_\fo$ over $\FOEncoding(\Sigma)$
and vice versa.
Observe that corresponding structures equally satisfy the theory axiomatization,
since they agree on all theory symbols,
and the satisfaction relation is defined identically.

\begin{claim}
	\label{thm:corresponding-structs-theory}
	If $M, M_\fo$ are corresponding structures, then 	
    $M \models \Gamma^\bg
		\iff M_\fo \models_\fo \Gamma^\bg$.
\end{claim}

\Proof{thm:corresponding-structs-theory}{
	This follows trivially, since
	the restriction of $M$ and $M_\fo$ to theory symbols (from $\Sigma^\bg$)
	results in identical structures, and
	for pure-theory formulas
	the satisfaction relations for \seplog{} and \fo{} coincide.
}

Having defined the \fo{} vocabulary and the correspondence between
\wsl{} and \fo{} structures,
we now continue with the translation of entailments modulo SID
which consists of two parts:
the translation of universal, conjunctive formulas,
which is the basic building block of the translation,
and the translation of systems of inductive definitions,
which builds upon the translation of universal, conjunctive formulas.

\paragraph{Translation of universal, conjunctive \seplog{} formulas}
Since we only consider determined-heap structures,
every atomic formula has a unique heaplet (if any)
that satisfies it.
This property is maintained through conjunction and universal quantification,
allowing us to map every universal conjunctive formula to its supporting heaplet.
Accordingly, in analogy to how an inductive predicate $P$
is captured by a pair $P_\fo, P_\eta$,
we translate every universal conjunctive  \seplog{} formula $\varphi$
into a pair of formulas $\varphi_\fo, \varphi_\eta$,
where $\varphi_\eta$ records the expected heaplet
of $\varphi$,
and $\varphi_\fo$ captures the tuples of elements
that satisfy $\varphi$,
using the heaplet-recording formulas
to express heap restrictions.
Intuitively, each occurrence of an inductive predicate $P$ in $\varphi$
is replaced by $P_\fo$ in $\varphi_\fo$,
each separating conjunction adds a restriction
that the expected heaplets of the two sub-formulas are disjoint,
and each conjunction and universal quantifier
add a restriction that the expected heaplets of the sub-formulas
are equal.
As for $\varphi_\eta$,
the encoding uses the free variable $\eVar$ to capture
the locations in the expected heaplet.
When $\varphi$ is an application of an inductive predicate $P$,
this is done using $P_\eta$,
and the other cases follow the semantics in the obvious way.

\begin{definition}
	Given a universal conjunctive formula \seplog{} $\varphi$
	over a vocabulary $\Sigma$,
	we define a transformation of $\varphi$
	into FOL as a pair of formulas
	$\varphi_\fo, \varphi_\eta$
	over $\FOEncoding(\Sigma)$,
	defined recursively in \Cref{tbl:fo-encoding}.
\end{definition}

\begin{table}[h]
	\footnotesize
	\centering
	\caption{Translation of universal conjunctive \seplog{} formula into \fol{}.}
	\label{tbl:fo-encoding}
	\begin{tabular}{| l | l | l |}
		\hline
		\multicolumn{1}{|c|}{$\varphi$} &
		\multicolumn{1}{c|}{$\varphi_\fo$} &
		\multicolumn{1}{c|}{$\varphi_\eta$}
		\\\hline
		$t_1 \bowtie t_2$ & $t_1 \bowtie t_2$ & $\false$
		\\\hline
		$Q(t_1, \dots, t_k)$ & $Q(t_1, \dots, t_k)$ & $\false$
		\\\hline
		$\neg Q(t_1, \dots, t_k)$ & $\neg Q(t_1, \dots, t_k)$ & $\false$
		\\\hline
		$\emp$ & $\true$ & $\false$
		\\\hline
		$P(t_1, \dots, t_k)$
		& $P_\fo(t_1, \dots, t_k)$
		& $P_\eta(t_1, \dots, t_k, \eVar)$
		\\\hline
		$t \points{t_1, \dots, t_n}$
		&
		$
		t \neq \nil \land \bigwedge_{i=1}^n t_i = m_i(t)
		$
		& $\eVar=t$
		\\\hline
		$\alpha \land \beta$
		& $\alpha_\fo \land \beta_\fo \land
		\forall \eVar. \alpha_\eta \liff \beta_\eta$
		& $\alpha_\eta$
		\\\hline
		$\alpha * \beta$
		& $\alpha_\fo \land \beta_\fo \land
		\neg \exists \eVar. \alpha_\eta \land \beta_\eta$
		& $\alpha_\eta \lor \beta_\eta$
		\\\hline
		$\forall x. \alpha$
		& $\parens{\forall x. \alpha_\fo} \land
		\forall x_1, x_2. \forall \eVar.
		\alpha_\eta \SingSubstitute{x}{x_1}
		\liff
		\alpha_\eta \SingSubstitute{x}{x_2}
		$
		& $\alpha_\eta$
		\\\hline
	\end{tabular}
\end{table}

To formalize the role of $\varphi_\eta$ as encoding the expected heaplet,
we use the following definition,
which defines the set of locations represented by
values of $\eVar$
in satisfying assignments of
$\varphi_\eta$.

\begin{definition}
	\label{def:corresponding-heaplet}
	Given a FO structure $M_\fo$, assignment $v$,
	and \seplog{} formula $\varphi$,
	we denote
	\[
	\bbrackets{\varphi}^{M_\fo}_v \defeq \midbraces{
		d \in \Dom(\fgSort)
	}{
		M_\fo, v\SingSubstitute{\eVar}{d} \models \varphi_\eta
	}.
	\]
\end{definition}

We formalize the aforementioned properties of the translation
with the following two claims.

\begin{claim}
	\label{thm:corresponding-struct-heaplet}
	For any universal conjunctive formula $\varphi$,
	corresponding models $M, M_\fo$,
	assignment $v$,
	and heaplet $\eta \subseteq \Dom(\fgSort)$,
	we have that
	$M,v,\eta \models \varphi
	\Rightarrow
	\eta = \bbrackets{\varphi}^{M_\fo}_v$.
\end{claim}

\Proof{thm:corresponding-struct-heaplet}{
	Structural induction over $\varphi$.
}

\begin{claim}
	\label{thm:sat-correspond}
	For any universal conjunctive formula,
	corresponding models $M, M_\fo$,
	and assignment $v$, we have that
	$M, v, \bbrackets{\varphi}^{M_\fo}_v \models
	\varphi
	\iff
	M_\fo, v \models \varphi_\fo
	$.
\end{claim}

\Proof{thm:sat-correspond}{
	Structural induction over $\varphi$.
}

\paragraph{Translation of systems of inductive definitions}
We now turn to encoding the \fp{} semantics
of systems of inductive definitions
into \fol{}.
As before,
we divide the encoding into two parts:
one that encodes the property that the interpretation
of the inductive predicates forms a fixpoint,
with respect to a SID,
and the other ensures that the supporting heaplets
of inductive predicates
match the expected heaplets of their definitions.

\begin{definition}
	\label{def:sys-encoding}
	For a SID $\sys$
	and inductive predicate $P$,
	let $\sys(P) = \exists \Vec{y}. \biglor_j \rho^P_j \parens{ \Vec x, \Vec y }$,
	we define the following axioms
	{\smaller\[
	\begin{array}{rcl}
	\sys_\fo(P)(\Vec x)
	&\defeq&
	\displaystyle
		P_\fo(\Vec{x}) \liff \exists \Vec{y}. \biglor_j \parens{ \rho^P_j }_\fo,
	\\
	\sys_\eta(P) (\Vec x)
	&\defeq&
	\displaystyle
	\forall \Vec y.
	\bigland_j \parens{
		\parens{\rho^P_j}_\fo
		\to
		\forall \eVar. P_\eta(\Vec x, \eVar) \liff \parens{\rho^P_j}_\eta
	}
	.
	\end{array}
	\]}
	and define the translation of the system as
	\[
	\sys_\fo \defeq \bigland_{P \in \Sigma^\fg} \forall \Vec x.
		\sys_\fo(P) \land \sys_\eta(P)
	.
	\]
\end{definition}

We formalize the correctness of the SID encoding
in the following claim.

\begin{claim}
	\label{thm:sid-corresponding-structures}
	Given a system $\sys$,
	and corresponding models $M, M_\fo$,
	\[
	M \in \Models_\fp(\sys) \iff M_\fo \models_\fo \sys_\fo.
	\]
\end{claim}

\Proof{thm:sid-corresponding-structures}{
	\
	\paragraph{Direction $\Rightarrow$}
	Let $M \in \Models_\fp(\sys)$,
	we need to prove that for any
	predicate
	$P \in \Sigma^\fg$,
	assignment $v$,
	and elements $d_1, \dots, d_k$
	the following holds:
	\[
	M_\fo, v' \models_\fo \sys_\fo(P) \land \sys_\eta(P),
	\]
	where $v' = v \SingSubstitute{\Vec x}{\Vec d}$.
	
	Let us first consider the case where
	$M_\fo, v' \not\models_\fo
	\exists \Vec y. \biglor_j \parens{ \rho^P_j }_\fo
	$.
	Thus, for any $j$, $\Vec d'$ elements
	$M_\fo, v''
	\not\models_\fo \parens{ \rho^P_j }_\fo$,
	where
	$v'' = v' \SingSubstitute{\Vec y}{\Vec d'} $.
	In particular,
	$M_\fo, v' \models_\fo \sys_\eta(P)$ trivially,
	since the antecedent in the inner conjunction never holds.

	Moreover, from \Cref{thm:sat-correspond}
	we get that
	$M, v'', \bbrackets{ \rho^P_j }^{M_\fo}_{v''} \not\models \rho^P_j$,
	and from \Cref{thm:corresponding-struct-heaplet}
	we get that for all $\eta$,
	$M, v'', \eta \not\models \rho^P_j$,
	and thus for all $\eta$
	$M, v', \eta \not\models \sys(P)$.
	Since $M \in \Models_\fp(\sys)$ we get,
	by definition \Cref{def:sid-sat}
	that for all $\eta$,
	$\tuple{d_1, \dots, d_k, \eta} \notin P^M$.
	
	Since $M$ and $M_\fo$ are corresponding we get that
	$\tuple{d_1, \dots, d_k} \notin P_\fo^{M_\fo}$
	and  in particular,
	$M, v' \not\models_\fo P_\fo(\Vec x)$.
	Therefore $M, v' \models_\fo \sys_\fo(P)$,
	and we get $M, v' \models_\fo \sys_\fo(P) \land \sys_\eta(P)$,
	as desired.
	
	\paragraph{Direction $\Leftarrow$}
	Let $M_\fo \models_\fo \sys_\fo$.
	We need to prove that for any
	predicate $P \in \Sigma^\fg$,
	assignment $v$,
	elements $d_1, \dots, d_k$
	and heaplet $\eta$,
	$\tuple{d_1, \dots, d_k, \eta} \in P^M$
	iff
	$M, v', \eta \models \sys(P)$
	where $v' = v \SingSubstitute{\Vec x}{\Vec d}$.
	Or in other words,
	iff there exists $j$ and elements $\Vec d'$
	s.t.
	$M, v'', \eta \models \rho^P_j$,
	where
	$v'' = v' \SingSubstitute{\Vec y}{\Vec d'}$.
	
	Since $M$ and $M_\fo$ are corresponding,
	$\tuple{d_1, \dots, d_k, \eta} \in P^M$
	iff
	$\tuple{d_1, \dots, d_k} \in P_\fo^{M_\fo}$
	and
	for all $d$,
	\[
	d \in \eta \iff \tuple{d_1, \dots, d_k, d} \in P_\eta^{M_\fo}.
	\]
	Since $M_\fo \models_\fo \sys_\fo$,
	and in particular
	$M_\fo, v' \models \sys_\fo(P)$,
	the elements
	$\tuple{d_1, \dots, d_k} \in P_\fo^{M_\fo}$
	iff
	there exists $j$ and $\Vec d'$
	s.t.
	$M_\fo, v'' \models_\fo \parens{\rho^P_j}_\fo$,
	where
	$v'' = v'  \SingSubstitute{\Vec y}{\Vec d'}$.
	For such $j$ and $\Vec d'$,
	we know that since
	$M_\fo, v' \models_\fo \sys_\eta(P)$,
	in particular
	\[
	M_\fo, v'' \models_\fo
		\parens{ \rho^P_j }_\fo
		\to
		\forall \eVar. P_\eta \parens{ \Vec x, \eVar } \iff \parens{ \rho^P_j }_\eta
		.
	\]
	From \Cref{thm:sat-correspond},
	$M_\fo, v'' \models_\fo \parens{ \rho^P_j }_\fo$
	iff
	$M, v'', \bbrackets{ \rho^P_j }^{M_\fo}_{v''} \models \rho^P_j$,
	and by definition
	\Cref{def:corresponding-heaplet}
	and
	$\forall \eVar. P_\eta \parens{ \Vec x, \eVar } \iff \parens{ \rho^P_j }_\eta$,
	iff
	$M, v'', \eta' \models \rho^P_j$
	where
	$\eta' = \midbraces{ d }{ \tuple{d_1, \dots, d_k, d} \in P_\eta^{M_\fo} }$.
	It is easy to see that since $M$ and $M_\fo$ are corresponding,
	it must be that $\eta' = \eta$.
	
	To recap,
	$\tuple{d_1, \dots, d_k, \eta} \in P^M$
	iff
	there exists $j$ and elements $\Vec d'$
	s.t.
	$M, v'', \eta \models \rho^P_j$,
	where
	$v'' = v \SingSubstitute{\Vec x}{\Vec d} \SingSubstitute{\Vec y}{\Vec d'}$,
	as desired.
}

We now arrive at the complete encoding of
entailments in \eprwsl{} modulo SIDs.
Given an entailment
$\varphi \sysentails_\wsl \exists \Vec u. \biglor_i \psi_i$
the basis of the encoding is to replace each of
$\sys, \varphi$ and $\psi_i$ with their \fo{} counterparts,
$\sys_\fo, \varphi_\fo$ and $\parens{\psi_i}_\fo$.
This is indeed necessary for guaranteeing
the correctness of the encoding,
but is not sufficient.
To see this,
consider the simpler case of
$\varphi \entails_\wsl \psi$,
where $\varphi$ and $\psi$
are both universal conjunctive \seplog{} formulas.
As exemplified by \Cref{thm:sat-correspond},
satisfaction of
$\psi_\fo$ by some structure $M_\fo$ and assignment $v$
corresponds to satisfaction of $\psi$
by $M, v$ and \emph{the expected heaplet} of $\psi$,
namely
$\bbrackets{\psi}^{M_\fo}_v$.
However, for the validity of the entailment
we must have
that for \emph{any} heaplet $\eta$,
if $M, v, \eta \models \varphi$
then $M, v, \eta \models \psi$.
Since $\varphi$ and $\psi$
are universal conjunctive \seplog{} formulas,
there is at one expected heaplet for each,
and this requirement translates to the condition
that for any structure and assignment that satisfy $\varphi$
with its expected heaplet,
they also satisfy $\psi$ with its expected heaplet,
\emph{and both expected heaplets are the same}:
$\forall \eVar. \varphi_\eta \liff \psi_\eta$.
The same principle generalizes to entailments
with consequents of the form
$\exists \Vec u. \biglor_i \psi_i$,
as shown in the following theorem.

\begin{theorem}[WSL/\fo{} Encoding]
		\label{thm:wsl-fo-encoding}
	A \wsl{} entailment
	$\varphi \sysentails_\wsl \exists \Vec u. \biglor_i \psi_i$
	is valid iff
	{\smaller\[
	\Gamma^\bg
	\models_\fo
		\parens{ \sys_\fo \land \varphi_\fo } \to
		\exists \Vec u. \biglor_i \parens{
		\parens{\psi_i}_\fo \land
		\forall \eVar. \varphi_\eta
			\liff \parens{\psi_i}_\eta
		},
	\]}
	or, equivalently, iff
	{\smaller\[
	\Gamma^\bg \cup \braces{
		\sys_\fo, \varphi_\fo,
		\forall \Vec{u}. \bigland_i \parens{
			\parens{\forall \eVar. \varphi_\eta
				\liff \parens{\psi_i}_\eta}
			\to
			\neg \parens{\psi_i}_\fo
		}
	}
	\]}
	is unsatisfiable
	(where $\Gamma^\bg$ is the first-order axiomatization of the theory).
\end{theorem}

\Proof{thm:wsl-fo-encoding}{
	\
	\paragraph{Direction $\Rightarrow$}
	Let us assume that the \wsl{} entailment is valid.
	Therefore,
	for every
	model $M$, assignment $v$ and heaplet $\eta$
	s.t.
	$M \in \Models_\fp(\sys)$
	satisfies the first-order axiomatization of the background theory
	$M \models \Gamma^\bg$,
	and
	$M, v, \eta \models \varphi$,
	then
	$M, v, \eta \models \exists \Vec u. \biglor_i \psi_i$.
	We need to prove that for any \fo{} model $M_\fo'$ and assignment $v$,
	if $M'_\fo, v \models_\fo \Gamma^\bg$
	then
	\[
	M'_\fo, v \models_\fo
	\parens{ \sys_\fo \land \varphi_\fo } \to
	\exists \Vec u. \biglor_i \parens{
		\parens{\psi_i}_\fo \land
		\forall \eVar. \varphi_\eta
		\liff \parens{\psi_i}_\eta
	}
	\]
	
	Let $M'_\fo$ be some \fo{} model s.t.
	$M'_\fo \models \Gamma^\bg$.
	
	If $M'_\fo, v \not\models_\fo \sys_\fo \land \varphi_\fo$,
	then we are done.
	
	Otherwise, by
	\Cref{thm:sat-correspond,thm:sid-corresponding-structures},
	for the corresponding model $M'$ it holds that
	$M' \in \Models_\fp(\sys)$
	and $M', v, \eta \models \varphi$,
	where
	$\eta = \bbrackets{ \varphi }^{M'_\fo}_v$.
	By our assumption,
	$M', v, \eta \models \exists \Vec u. \biglor_i \psi_i$.
	In particular, there exist elements $\Vec d$ and some $i$
	s.t.
	$M', v', \eta \models \psi_i$,
	where
	$v' = v \SingSubstitute{\Vec u}{\Vec d}$.
	By \Cref{thm:sat-correspond,thm:corresponding-struct-heaplet},
	$M'_\fo, v' \models_\fo \parens{ \psi_i }_\fo$
	and
	$\eta = \bbrackets{ \psi_i }^{M'_\fo}_{v'}$.
	Since
	$\eta = \bbrackets{ \varphi }^{M'_\fo}_v$,
	by \Cref{def:corresponding-heaplet},
	$M'_\fo, v' \models_\fo \forall \eVar. \varphi_\eta \liff \parens{ \psi_i }_\eta$,
	and thus
	\[
	M'_\fo, v \models_\fo
	\parens{ \sys_\fo \land \varphi_\fo } \to
	\exists \Vec u. \biglor_i \parens{
		\parens{\psi_i}_\fo \land
		\forall \eVar. \varphi_\eta
		\liff \parens{\psi_i}_\eta
	},
	\]
	as desired.
	
	\paragraph{Direction $\Leftarrow$}
	Note that the proof for $\Rightarrow$ hinges on
	\Cref{thm:sat-correspond,thm:sid-corresponding-structures},
	which are ``$\iff$'' claims,
	and thus the proof for $\Leftarrow$
	works much in the same way,
	where implication arrows are reversed.
}

We can thus conclude that for every SID $\sys$ ,
finite set of \eprwsl{} formulas $\Gamma$
and \eprwsl{} formula $\psi$,
the entailment $\Gamma \sysentails_\wsl \psi$
is reducible to a \fol{} formula $\alpha_\fo$
such that the entailment is valid
(modulo theory $\bg$)
iff
$\Gamma^\bg \models_\fo \alpha_\fo$.
\neta{clunky?}
In case $\sys$ and all formulas are theory-free
we can simply replace $\Gamma^\bg$ with $\emptyset$.
We now reach the culmination of the \fo{} encoding.
By having an equivalence between valid entailments in \eprwsl{}
and entailments in \fol{},
we can carry key properties of \fol{}
over to \wsl{}.
In particular, since $\Gamma^\bg$
is a recursive set of first-order sentences,
\fol{} has a complete proof system for
checking $\Gamma^\bg \models_\fo \alpha_\fo$.
From this follows the completeness of \eprwsl{}:

\begin{theorem}[\eprwsl{} Completeness]
	\label{thm:wsl-complete}
	\eprwsl{} admits a complete proof system
	for entailments modulo systems of inductive definitions.
\end{theorem}

\Proof{thm:wsl-complete}{
	As explained above,
	any entailment
	$\Gamma \sysentails_\wsl \psi$
	in \eprwsl{}
	is equivalent to a finite set of entailments
	$\varphi_j \sysentails_\wsl \psi$,
	where each $\varphi_j$ is a universal, conjunctive formula.
	From \Cref{thm:wsl-fo-encoding},
	we know that checking the validity of each
	$\varphi_j \sysentails_\wsl \psi$
	is equivalent to checking the unsatisfiability of a set of formulas in \fol{}.
	Since \fol{} admits a complete proof mechanism,
	\eprwsl{} also admits a complete proof mechanism.
}

This gives a complete but perhaps unnatural proof mechanism
for \eprwsl{}:
translation into \fol{} and then using \fol{} proof systems.
In the next section we study a more natural proof mechanism
for \wsl{} and \seplog{}: the ubiquitous fold/unfold mechanism.

Note that this completeness result does not imply decidability,
and in fact \eprwsl{} is not decidable.
This can be shown by encoding a tiling problem
as a system of inductive definitions
(e.g., predicates representing
whether a coordinate in the plane is covered by some tile)
and constructing an entailment that is valid exactly when there is no way to tile
the top-right quadrant of the plane.

\begin{theorem}[\eprwsl{} Undecidability]
	\label{thm:wsl-undecidable}
	The validity problem of
	\eprwsl{}
	entailments modulo systems of inductive definitions
	is undecidable.
\end{theorem}

\Proof{thm:wsl-undecidable}{
	Without restrictions on the background theory,
	decidability can trivially come from the theory,
	as it can be undecidable itself or undecidable
	in conjunction with quantified formulas containing uninterpreted relations
	(as allowed in \eprsl{}).
	We shall prove that even the theory-free fragment of \eprwsl{}
	is undecidable by giving a reduction from tilings problems
	to entailments in \eprwsl{}.

	The high-level idea is to use the system of inductive definitions
	to define a predicates for
	\begin{enumerate*}[(a)]
		\item whether a coordinate on the plane is covered
		by some tile;
\item whether a coordinate is covered by two
		different tiles;
\item whether a coordinate is covered by a tile
		conflicting with its neighbors; and
\item predicates to define a ``successor'' function ($s$),
		for encoding the coordinate system.
	\end{enumerate*}

	Given a set of tiles $T_1, \dots, T_n$
	we fix a vocabulary
	\[
	\Sigma^\fg = \braces{
		\begin{array}{c}	
			c,
			\\
			P_1(\cdot, \cdot), \dots, P_n(\cdot, \cdot),
			\\
			P_s(\cdot, \cdot), P_{s-tot}(\cdot), P_{s-nonfunc}(\cdot),
			\\
			\covered(\cdot,\cdot),
			\conflict(\cdot,\cdot),
			\unmatchTop(\cdot,\cdot),
			\unmatchRight(\cdot,\cdot)
		\end{array}
	},
	\]
	and give the following inductive definitions:
	\begin{enumerate}
		\item
		We define the predicates $P_1, \dots, P_n$ and $P_s$ as completely uninterpreted:
		\[
		P_i(x,y) \coloneq P_i(x,y),
		\]
		and
		\[
		P_s(x,y) \coloneq P_s(x,y).
		\]

		\item
		We use $P_{s-tot}(x)$ to encode when $P_s$ is defined for $x$:
		\[
		P_{s-tot}(x) \coloneq \exists y. P_s(x,y).
		\]

		\item
		We use $P_{s-nonfunc}(x)$ to encode that $P_s$ has non-univalent
		definition at $x$:
		\[
		P_{s-nonfunc}(x) \coloneq \exists y, z. P_s(x,y) \land P_s(x,z) \land y \neq z.
		\]

		\item
		We use $\covered(x,y)$ to encode when some tile is covering the $(x,y)$ coordinate:
		\[
		\covered(x,y) \coloneq P_1(x,y) \,\mid\, \dots \,\mid\, P_n(x,y).
		\]

		\item
		We use $\conflict(x,y)$ to encode two conflicting tiles at the $(x,y)$ coordinate:
		\[
		\conflict(x,y) \coloneq P_1(x,y) \land P_2(x,y)
			\,\mid\, P_1(x,y) \land P_3(x,y)
			\,\mid\, \dots \,\mid\, P_{n-1}(x,y) \land P_n(x,y).
		\]

		\item
		Finally, we use $\unmatchTop(x,y)$ and
		$\unmatchRight(x,y)$ to encode a coordinate with
		unmatched tiles to the top or to the right
		(respectively). For example, if $T_i$'s top color
		does not match $T_j$'s bottom color,
		$\unmatchTop(x,y)$ will include a case $\exists z.
		P_s(y,z) \land P_i(x,y) \land P_j(x,z)$.
	\end{enumerate}

	We use the following entailment to encode that
	there exists no possible tiling of the plane:
	\[
	\forall x,y. P_{s-tot}(x) \land \covered(x,y)
	\entails
	\exists x,y.
	\parens{
		\begin{array}{rl}
			& P_{s-nonfunc}(x) \\
			\lor & \conflict(x,y) \\
			\lor & \unmatchTop(x,y) \\
			\lor & \unmatchRight(x,y)
		\end{array}
	}.
	\]
}   

\section{Fold/Unfold Proof Mechanisms: Soundness and Completeness}
\label{sec:unfolding}

Equipped with the tools of \fol{},
we now turn to formalizing fold/unfold proof mechanisms
of separation logic,
and discussing their soundness and completeness in regards to both
standard Separation Logic (\seplog)
and Weak Separation Logic (\wsl).
We side-step the need for quantifier instantiations
in \seplog{} proof mechanisms by restricting ourselves
to theory-free SIDs
and theory-free, quantifier-free formulas.
Similarly to entailments in \eprwsl{}, we need only consider the case
$\varphi \sysentails_\Logic \biglor_i \psi_i$
where $\varphi,\psi_i$ are conjunctive formulas,
since we can always replace a finite set of assumptions
by a single assumption,
and we can then eliminate disjunctions in the antecedent
by splitting the entailment.

We show that though fold/unfold is sound for both \seplog{} 
and \wsl{},
and expectedly incomplete for \seplog,
surprisingly, it is complete for \wsl{}. 
In this sense, \wsl{} is the appropriate logic to consider
when working with fold/unfold proof mechanisms.

We start by deriving a formal definition of an \emph{abstract}
fold/unfold proof mechanism,
motivated by real-world tools for \seplog{},
but glossing over irrelevant details and minutia.

\subsection{Abstract Fold/Unfold Proof Mechanism}
The heart of a fold/unfold proof mechanism
lies in its two namesake transformations:
folding, which replaces a formula that matches
one of the cases in a definition
of an inductive predicate with its application,
and unfolding, which expands an inductive predicate 
into its definition.
Though in practice these transformations happen ``in-place'',
in the abstract they can be thought of as two kinds of axioms added
to the antecedent of the entailment.
These axioms are thus the design principle behind
fold/unfold proof mechanisms,
whereas any specific choice of when to 
fold or unfold is an implementation detail.
In the sequel we therefore define 
an abstract fold/unfold proof mechanism
as a recursively enumerable set of finite sets of such axioms,
where for each set of axioms,
we check the validity of entailment,
\emph{modulo these axioms}.
A proof is found if one of these entailments turns out to be valid.

We start with a formal definition of fold/unfold axioms.

\begin{definition}[Fold/unfold axioms]
	\label{def:fold-unfold-axioms}
	Let $\sys$ be a SID, 
	$P \colon \sort_1 \tdots \sort_k$
	an inductive predicate,
	and 
	$\sys(P) = \exists y_1, \dots, y_\ell. \biglor_j \rho^P_j$.
	Given 
	terms $\Vec{t} = \tuple{t_1, \dots, t_k}$
	and $\Vec t' = \tuple{t'_1, \dots, t'_\ell}$,
	a \emph{fold axiom} of $\sys(P)$
	with terms $\Vec t, \Vec t'$
	is the \seplog{} formula
	{\smaller\[
	\fold(\sys(P), \Vec t, \Vec t') \defeq
		\bigland_j \parens{
		\rho^P_j\brackets{
			\Substitute{\Vec x}{\Vec t},
			\Substitute{\Vec y}{\Vec t'}
		} \to P(\Vec t)
	}.
	\]}
	
	Given terms $\Vec t$ and \emph{fresh constants}
	$\Vec c = \tuple{c_1, \dots, c_\ell}$
	an \emph{unfold axiom} of $\sys(P)$ with terms 
	$\Vec{t}, \Vec c$
	is the \seplog{} formula	
	{\smaller \[
	\unfold(\sys(P), \Vec t, \Vec c)
	\defeq
	P(\Vec t) \to
	\biglor_j \rho^P_j\brackets{
		\Substitute{\Vec x}{\Vec t},
		\Substitute{\Vec y}{\Vec c}
	}.
	\]}
\end{definition}

Note that the terms $\Vec t, \Vec t'$ used in fold/unfold axioms
can themselves include fresh constants.
Importantly, even though SIDs may include existential quantifiers, 
observe that all fold/unfold axioms are quantifier-free (and theory-free).

\begin{note}
We extend the satisfaction relation
(of \Cref{def:sl-sat})
to the $\to$ connective in the obvious way.
\end{note}

The small-model property of theory-free, quantifier-free \seplog{} formulas
is maintained in the presence of the $\to$ connective.
This means that \Cref{thm:sl-wsl-equi-valid,thm:qf-sl-decidable}
continue to hold
when the finite set of assumptions in entailments
includes fold/unfold axioms.
That is, $U \cup \Gamma \entails_\seplog \psi$
is valid iff
$U \cup \Gamma \entails_\wsl \psi$
is valid,
and checking validity for both kinds of entailments is decidable.
Decidability justifies using such entailments 
as the base case of fold/unfold proof mechanisms,
which iteratively produce finite sets of fold/unfold axioms
and add them as assumptions. \neta{cut: for entailments.}
Interestingly, 
entailments with finitely many fold/unfold
axioms are equi-valid in \seplog{} and \wsl{},
even though they are not
equi-valid modulo SIDs. 
This discrepancy is manifested in
the (in)completeness of fold/unfold mechanisms: 
as we show next, 
this procedure is complete for \wsl{}, but not for \seplog{}.

We now give a formal definition of 
an abstract fold/unfold proof mechanism.

\begin{definition}[Abstract fold/unfold proof mechanism]
	An abstract fold/unfold proof mechanism
	is a procedure that enumerates finite sets of 
	fold/unfold axioms $U_1, U_2, \dots$,
	and for each set $U_i$ checks the validity of the entailment
	$\Gamma \cup U_i \entails_\seplog \psi$.
	The procedure returns ``valid'' when any such entailment is valid,
	and must be exhaustive,
	in the sense that for every finite set $U$ of fold/unfold axioms,
	there exists some $i$ s.t. $U \subseteq U_i$.
\end{definition}

Typically, real-world fold/unfold proof mechanisms
produce the sets of axioms $U_1, U_2, \dots$
by enumerating individual axioms
and accumulating them.
In this case, exhaustiveness can be achieved
by ensuring that all fold/unfold axioms
ultimately occur in the enumeration.
Note that an abstract fold/unfold proof mechanism uses validity checks in SL;
however, this may be equivalently done by validity checks in WSL since, 
as explained above, 
the two logics coincide for the entailments obtained 
by the addition of fold/unfold axioms.

In order to analyze the abstract fold/unfold proof mechanism 
we first give a formal definition for the soundness and completeness
of a proof mechanism,
and then establish soundness and (in)completeness of
abstract fold/unfold for quantifier-free \seplog{} and \wsl{}.

\begin{definition}[Soundness and completeness]
	\neta{changed}
	A proof mechanism is sound 
	for a fragment of a logic
	$\Logic \in \braces{\seplog, \wsl}$
	if whenever the mechanism returns ``valid''
	for an entailment 
	$\Gamma \sysentails_\Logic \psi$
	within the fragment,
	then $\Gamma \sysentails_\Logic \psi$ is indeed valid.
A proof mechanism is complete if whenever 
	$\Gamma \sysentails_\Logic \psi$ is valid
	it halts and returns ``valid''.
\end{definition}

\neta{cut:
Note that while we define two kinds of entailments
for each logic
(plain or modulo SIDs),
soundness and completeness of proof mechanisms
are always defined w.r.t. entailment modulo SIDs.
}

An immediate consequence of the definition
of fold/unfold axioms
is that any \fp{} structure of $\sys$
and any heaplet
satisfy
all of them.
From this follows the soundness of
fold/unfold for both \seplog{},
where only \lfp{} structures are considered,
and \wsl{},
where any \fp{} structure is considered.

\begin{theorem}[Soundness]
	\label{thm:fold-unfold-sound}
	Any fold/unfold proof mechanism is sound for both
	\seplog{} and \wsl{}.
\end{theorem}

\Proof{thm:fold-unfold-sound}{
	Given a set $U$ of fold/unfold axioms
	for $\sys$,
	for any $M \in \Models_\fp(\sys)$,
	assignment $v$
	and heaplet $\eta$,
	we have that
	$M, v, \eta \models U$.
	Therefore,
	given a set of formulas $\Gamma$,
	if $\Gamma \cup U \entails_\Logic \psi$ is valid,
	then $\Gamma \sysentails_\Logic \psi$ is valid,
	where $\Logic \in \braces{\seplog, \wsl}$.
}

As opposed to soundness,
the gap between \seplog{} and \wsl{}
pokes its head when completeness is considered.
Namely, \seplog{} with inductive definitions is known 
to have no complete proof mechanism,
even when only theory-free, quantifier-free formulas are considered~\cite{seplog-incomplete}.
In contrast, we show that any abstract fold/unfold proof mechanism
is complete for the theory-free, quantifier-free fragment of \wsl{} with inductive definitions.

\begin{theorem}[Incompleteness for \seplog{}~\cite{seplog-incomplete}]
	\label{thm:sl-incomplete}
	\neta{changed}
	Any abstract fold/unfold proof mechanism is incomplete 
	for the theory-free, quantifier-free fragment of \seplog{}
	with theory-free SIDs.
\end{theorem}

\Proof{thm:sl-incomplete}{
	As the theory-free, quantifier-free fragment of \seplog{}
	with theory-free SIDs has no complete proof system
	(see, e.g.,~\cite{seplog-incomplete}),
	the abstract fold/unfold proof mechanism 
	is, in particular, not complete for this fragment.
}

\begin{theorem}[Completeness for \wsl{}]
	\label{thm:wsl-fold-unfold-complete}
	All abstract fold/unfold proof mechanisms are complete 
	for the theory-free, quantifier-free fragment of \wsl{}
	with theory-free SIDs.
\end{theorem}

In the sequel we prove \Cref{thm:wsl-fold-unfold-complete}
by extending the encoding of \Cref{sec:fo-encoding}
to relate fold/unfold axioms
and quantifier instantiations of $\sys_\fo$ axioms,
laying the ground to using Herbrand's theorem on the \fol{} side,
and carrying the result back over to \wsl{}.

\subsection{Fold/Unfold Axioms and Quantifier Instantiations}
\label{sec:unfolding-fo-encoding}

Before delving into the translation of fold/unfold axioms to \fol{}
let us first recall the \fo{} axioms we defined for each 
inductive predicate $P$ (\Cref{def:sys-encoding}):
{\smaller\[
	\sys_\fo(P)(\Vec x)
	\defeq
	P_\fo(\Vec{x}) \liff \exists \Vec{y}. \biglor_j \parens{ \rho^P_j }_\fo
	\qquad\qquad
	\sys_\eta(P) (\Vec x)
	\defeq
	\forall \Vec y. 
	\bigland_j \parens{
		\parens{\rho^P_j}_\fo
		\to
		\forall \eVar. P_\eta(\Vec x, \eVar) \liff \parens{\rho^P_j}_\eta
	}
\]}

These can be equivalently expressed by the following pair of axioms:

{\smaller\[
\begin{array}{rcl}
	\alpha_P(\Vec x) & \defeq&
	\displaystyle
		P_\fo(\Vec x) \to \exists \Vec{y}.
			\biglor_j 
				\parens{
					\parens{ \rho^P_j }_\fo
					\land 
					\parens {
						\forall \eVar. P_\eta(\Vec x, \eVar) \liff \parens{\rho^P_j}_\eta
					}
				},
		\\
	\beta_P(\Vec x) &\defeq&
	\displaystyle
	\forall \Vec y. 
	\bigland_j \parens{
		\parens{\rho^P_j}_\fo
		\to
		P_\fo(\Vec x) \land
		\forall \eVar. P_\eta(\Vec x, \eVar) \liff \parens{\rho^P_j}_\eta
	}
\end{array}
\]}

\neta{maybe we want a shorthand notation for the ``same heap'' sub-formula}

We now define the \fo{} encoding of fold/unfold axioms
and observe that it essentially produces quantifier instantiations
for the above axioms,
where $\alpha_P$ matches $\unfold_\fo$
and $\beta_P$ matches $\fold_\fo$.

\begin{definition}[Fold/unfold encoding]
	Given SID $\sys$, predicate $P$ and terms $\Vec t, \Vec t'$,
	we define the \fo{} encoding of the fold axiom
	$\fold \parens{ \sys(P), \Vec t, \Vec t' }$
	as follows
	{\smaller\[
	\fold_\fo \parens{ \sys(P), \Vec t, \Vec t' } \defeq
	\displaystyle
	\bigland_j \parens{
		\parens{\rho^P_j}_\fo \brackets{
			\Substitute{\Vec x}{\Vec t},
			\Substitute{\Vec y}{\Vec t'}
		}
		\to
		P_\fo(\Vec t) \land
		\forall \eVar. P_\eta(\Vec t, \eVar) \liff \parens{\rho^P_j}_\eta\brackets{
			\Substitute{\Vec x}{\Vec t},
			\Substitute{\Vec y}{\Vec t'}
		}
	},
	\]}
	given terms $\Vec t$ and fresh constants $\Vec c$,
	we define the \fo{} encoding of the unfold axiom
	as
	{\smaller\[
	\unfold_\fo \parens{ \sys(P), \Vec t, \Vec c } \defeq
	\displaystyle
	P_\fo(\Vec x) \to 
	\parens{
		\biglor_j 
		\parens{
			\parens{ \rho^P_j }_\fo \brackets{ 
				\Substitute{\Vec x}{\Vec t},
				\Substitute{\Vec y}{\Vec c}
			}
			\land \parens {
				\forall \eVar. P_\eta(\Vec t, \eVar) \liff 
				\parens{\rho^P_j}_\eta \brackets{ 
					\Substitute{\Vec x}{\Vec t},
					\Substitute{\Vec y}{\Vec c}
				}
			}
		}
	},
	\]}
	and finally we lift these encodings to a set
	$U$ of fold/unfold axioms,
	defining its encoding $U_\fo$
	as the result of replacing each fold/unfold axiom in $U$
	with its corresponding
	$\fold_\fo$ / $\unfold_\fo$.
\end{definition}

The following claim formalizes 
that these definitions
faithfully encode 
the fold/unfold axioms.

\begin{claim}
	\label{thm:fold-unfold-encoding}
	Given corresponding structures $M, M_\fo$,
	assignment $v$,
	and set $U$ of fold/unfold axioms,
	we have that
	$M_\fo, v \models_\fo U_\fo$
	iff 
	$M, v, \eta \models U$
	for every heaplet $\eta$.
\end{claim}

\Proof{thm:fold-unfold-encoding}{
	For a single $\fold$ or $\unfold$ axiom,
	the proof works in the same way as
	\Cref{thm:sat-correspond,thm:sid-corresponding-structures},
	and it is trivially lifted to any set of axioms $U$.
}

\subsection{Bringing It Full Circle}
We finally have in place all the pieces needed
to show that an abstract fold/unfold proof mechanism
is complete for 
checking the validity of
theory-free, quantifier-free \wsl{} entailments
modulo theory-free SIDs.
The proof leverages the \fo{} encoding of entailments in \wsl{} 
with SIDs or with fold/unfold axioms 
to reduce the completeness claim 
to a claim about entailments in \fol{}, 
where we can apply \fol{} results such as Herbrand's theorem and compactness.
The structure of the proof is shown in
\Cref{fig:fold-unfold-proof}.
Considering theory-free, quantifier-free formulas 
and theory-free SIDs, the proof works by using a chain of 
three equivalences to conclude the fourth: 
\begin{enumerate*}[(i)]
	\item
	an entailment modulo SID is valid in \wsl{}
	iff its encoding is valid in \fol{} (follows from correctness of the encoding);
\item
	the encoding of an entailment modulo SID
	is valid in \fol{}
	iff
	there exists some finite set of quantifier instantiations of the system,
	modulo which the encoding of the entailment is valid 
	(follows from Herbrand's theorem and compactness in \fol{});
\item
	the encoding of the entailment modulo 
	a finite set of quantifier instantiations of the system is valid
	(in \fol{})
	iff
	the entailment modulo a corresponding finite set of fold/unfold axioms
	is valid in \wsl{} (follows from the correctness of the encoding); and, transitively,
\item
	an entailment modulo SID is valid in \wsl{}
	iff 
	there exists a finite set of fold/unfold axioms
	modulo which the entailment is valid.
\end{enumerate*}
Thus, an exhaustive set of fold/unfold axioms
forms a complete proof mechanism for \wsl{}.
This line of reasoning is formalized in the following claims.

\begin{figure}
	\centering
	\scalebox{0.95}{
	\begin{tikzpicture}[
		x=1.35cm,
		y=0.8cm,
		quad/.style={
			draw, 
			minimum width=3.25cm,
			minimum height=1cm,
			text width=3.25cm,
			align=center,
			inner sep=1.5pt,
		},
		iff/.style={
			double,
			line width=1pt,          double distance=2.4pt,     arrows={Implies}-{Implies}
		}
		]

		\node (wsl) at (0, 4.25) {\wsl{}};
		\node (fol)  at  (4, 4.25) {\fol{}};
		\node (system) at (-2.5, 3) {SID $\sys$};
		\node (fold-unfold) at (-2.5, 0) {Fold/unfold};
		
		\node[quad] (A) at (0, 3) {$\varphi \sysentails_\wsl \psi$};
		\node[quad] (B) at (4, 3) 
		{$\braces{\sys_\fo , \varphi_\fo } \models_\fo \psi_\fo$};
		\node[quad] (C) at (0, 0) 
		{$U \cup \braces \varphi \entails_\wsl \psi$};
		\node[quad] (D) at (4, 0) 
		{$U_\fo \cup \braces {\varphi_\fo} \models_\fo \psi_\fo$};
		
\draw[dashed, thick] (2, -1.5) -- (2, 4.5); \draw[dashed, thick] (-3, 1.5) -- (7, 1.5); 

\draw[iff] (A) -- 
		node[midway, above=10pt, fill=white, inner sep=1pt] 
		{\Cref{thm:wsl-fo-encoding}} 
		(B);
		\draw[iff] (C) -- 
		node[midway, below=10pt, fill=white, inner sep=1pt] 
		{\Cref{thm:fold-unfold-encoding}} 
		(D);
		
\draw[iff,
			dash pattern=on 8.5pt off 4pt
		] (A) -- 
		node[midway, left=10pt, yshift=8pt, fill=white, inner sep=1pt] 
		{\Cref{thm:wsl-fold-unfold}} 
		(C);
		\draw[iff] (B) -- 
		node[midway, right=10pt, yshift=8pt, fill=white, inner sep=1pt] 
		{\Cref{thm:fol-fold-unfold}} 
		(D);
	\end{tikzpicture}
	}
	\caption{Abstract fold/unfold is a complete proof mechanism
		for \wsl{}: proof structure.
	}
	\label{fig:fold-unfold-proof}
\end{figure}
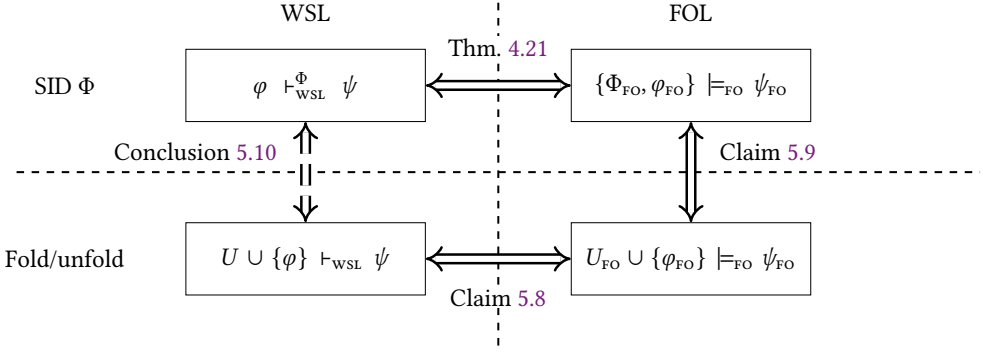 
\begin{claim}
	\label{thm:fol-fold-unfold}
	For a theory-free SID $\sys$,
	and a pair of
	theory-free, quantifier-free formulas $\varphi, \psi$,
	we have that
	$\braces{\sys_\fo , \varphi_\fo }\models_\fo \psi_\fo$
	iff
	there exists a finite set $U$
	of fold/unfold axioms for $\sys$
	s.t.
	$U_\fo \cup \braces{\varphi_\fo} \models_\fo \psi_\fo$.
\end{claim}

\ifproofs\else
\begin{proof}[Proof sketch]
	Observe that for theory-free formulas and SIDs,
	all applications of functions that appear in
	$\sys_\fo, \varphi_\fo$ and $\psi_\fo$
	are of the form $m_i(t) = t'$,
	where $t'$ has no function applications.
	Thus, when considering the set of all ground instances
	of $\sys_\fo, \varphi_\fo, \psi_\fo$,
	per Herbrand's theorem,
	we need only consider ground terms with no function applications.
	An implication of Herbrand's theorem is that 
	partially-ground instantiations,
	where an inner part of the formula stays universally quantified,
	is equi-satisfiable to the complete set of instantiations.
	Specifically, not instantiating the ``same heap'' condition 
	in the encoding
	of the SID and fold/unfold axioms
	does not detract from the result of Herbrand's theorem.
	Therefore, the set of partial instantiations to consider
	correspond to the encoding of all fold/unfold axioms,
	and from compactness, we have a finite set of such instantiations,
	or equivalently fold/unfold axioms,
	to consider.
\end{proof}
\fi

\Proof{thm:fol-fold-unfold}{
	First recall that
	$\sys_\fo = 
	\bigland_{P \in \Sigma^\fg} \forall \Vec x.
	\sys_\fo(P) \land \sys_\eta(P)
	$,
	or, equivalently,
	\[\sys_\fo = \bigland_P \parens{ 
		\parens{ \forall \Vec x. \alpha_P(\Vec x) }
		\land
		\parens{ \forall \Vec x. \beta_P(\Vec x) }
	}.
	\]
	This can be further transformed to the following
	equisatisfiable
	set of formulas:
	\[
	\sys_\fo \equiv 
	\Gamma_1 \defeq
	\bigcup_P
	\braces{ 
		\forall \Vec x. \exists \Vec y. \unfold_\fo \parens{
			\sys(P), \Vec x, \Vec y
		}
	}
	\cup
	\braces{ 
		\forall \Vec x, \Vec y. \fold_\fo \parens{
			\sys(P), \Vec x, \Vec y
		}
	}
	.
	\]
	Thus, we can prove \Cref{thm:fol-fold-unfold}
	by equivalently prove that the set of formulas
	\[
	\Gamma_2 \defeq \Gamma_1 \cup \braces{\varphi_\fo, \neg \psi_\fo} \cup
	\]
	is unsatisfiable.
	
	By Skolem's theorem, $\Gamma_2$ is unsatisfiable
	iff the set
	\[
	\Gamma_3 \defeq
	\braces{\varphi_\fo, \neg \psi_\fo} \cup
	\bigcup_P
	\braces{ 
		\forall \Vec x. \unfold_\fo \parens{
			\sys(P), \Vec x, \Vec {f_y}(\Vec x)
		}
	}
	\cup
	\braces{ 
		\forall \Vec x, \Vec y. \fold_\fo \parens{
			\sys(P), \Vec x, \Vec y
		}
	}
	\]
	is unsatisfiable,
	where $\Vec{f_y}$ are fresh Skolem function symbols.
	
	We denote 
	$\fold'_\fo$ and $\unfold'_\fo$
	the formulas resulting from the deleting the $\forall \eVar$ quantifier
	from $\fold_\fo, \unfold_\fo$ (respectively),
	leaving $\eVar$ free,
	and by using logical identifies we get 
	that the set $\Gamma_3$ is unsatisfiable
	iff
	\[
	\Gamma_4 \defeq \braces{\varphi_\fo, \neg \psi_\fo} \cup
	\bigcup_P
	\braces{ 
		\forall \Vec x, \eVar. \unfold'_\fo \parens{
			\sys(P), \Vec x, \Vec {f_y}(\Vec x), \eVar
		}
	}
	\cup
	\braces{ 
		\forall \Vec x, \Vec y, \eVar. \fold'_\fo \parens{
			\sys(P), \Vec x, \Vec y, \eVar
		}
	}
	\]
	is unsatisfiable.
	
	By Herbrand's theorem, 
	$\Gamma_4$ is unsatisfiable iff the set
	\[
	\Gamma_5 \defeq \braces{\varphi_\fo, \neg \psi_\fo} \cup
	\bigcup_{P, \Vec t, \Vec t', t^* \text{ are terms from } \Sigma^\fg_\fo }
	\braces{ 
		\unfold'_\fo \parens{
			\sys(P), \Vec t, \Vec {f_y}(\Vec t), t^*
		}
	}
	\cup
	\braces{ 
		\fold'_\fo \parens{
			\sys(P), \Vec t, \Vec t', t^*
		}
	}
	\]
	is unsatisfiable.
	
	We can replace the terms 
	$\Vec{f_y}(\Vec t)$
	by fresh constants $\Vec c_{\Vec t}$
	to get the equi-satisfiable set
	\[
	\Gamma_6 \defeq \braces{\varphi_\fo, \neg \psi_\fo} \cup
	\bigcup_{P, \Vec t, \Vec t', t^* \text{ are terms from } \Sigma^\fg_\fo }
	\braces{ 
		\unfold'_\fo \parens{
			\sys(P), \Vec t, \Vec c_{\Vec t}, t^*
		}
	}
	\cup
	\braces{ 
		\fold'_\fo \parens{
			\sys(P), \Vec t, \Vec t', t^*
		}
	}.
	\]
	
	By applying Herbrand's theorem in the inverse direction,
	and the logical identities relating $\fold_\fo$ and $\unfold_\fo$
	to $\fold'_\fo$ and $\unfold'_\fo$,
	we get that $\Gamma_6$ is unsatisfiable 
	iff the set
	\[
	\Gamma_7 \defeq \braces{\varphi_\fo, \neg \psi_\fo} \cup
	\bigcup_{P, \Vec t, \Vec t'* \text{ are terms from } \Sigma^\fg_\fo }
	\braces{ 
		\unfold_\fo \parens{
			\sys(P), \Vec t, \Vec c_{\Vec t}
		}
	}
	\cup
	\braces{ 
		\fold_\fo \parens{
			\sys(P), \Vec t, \Vec t'
		}
	}.
	\]
	is unsatisfiable.
	
	We observe 
	that for theory-free formulas and SIDs,
	all applications of functions that appear in
	$\sys_\fo, \varphi_\fo$ and $\psi_\fo$
	are of the form $m_i(t) = t'$,
	where $t'$ has no function applications.
	Thus, when considering the set of all ground instances
	of in $\Gamma_7$,
	we need only consider ground terms with no function applications,
	i.e.,
	the terms $\Vec t$ contain only symbols from the
	original, $\Sigma^\fg$ vocabulary. I.e., $\Gamma_7$
	is equisatisfiable to the set
	\[
	\Gamma_8 \defeq \braces{\varphi_\fo, \neg \psi_\fo} \cup
	\bigcup_{P, \Vec t, \Vec t' \text{ are terms from } \Sigma^\fg }
	\braces{ 
		\unfold_\fo \parens{
			\sys(P), \Vec t, \Vec c_{\Vec t}
		}
	}
	\cup
	\braces{ 
		\fold_\fo \parens{
			\sys(P), \Vec t, \Vec t'
		}
	}.
	\]
	
	By compactness, the set $\Gamma_8$ is unsatisfiable
	iff there exist an unsatisfiable finite subset $\Gamma_9 \subset \Gamma_8$.
	In other words, iff there exist a finite set $U$ of fold/unfold axioms
	for $\sys$ s.t.
	\[
	U_\fo \cup \braces{\varphi_\fo, \neg \psi_\fo}
	\]
	is unsatisfiable,
	iff
	$U_\fo \cup \braces \varphi_\fo \models_\fo \psi_\fo$,
	as desired.
}

\begin{conclusion}
	\label{thm:wsl-fold-unfold}
	\neta{we must shorten}
	For a theory-free SID $\sys$
	and theory-free, quantifier-free formulas
	$\varphi, \psi$,
	the entailment
	$\varphi \sysentails_\wsl \psi$ is valid
	iff
	there is a finite set of fold/unfold axioms $U$
	s.t.
	$U \cup \braces\varphi \entails_\wsl \psi$ is valid.
\end{conclusion}

Recall that plain entailment of theory-free, quantifier-free formulas
coincide in \seplog{} and \wsl{}, i.e.,
the entailment
$U \cup \braces\varphi \entails_\wsl \psi$
of \Cref{thm:wsl-fold-unfold}
coincides with
$U \cup \braces\varphi \entails_\seplog \psi$,
which is the check used in an abstract fold/unfold mechanism.
This completes the proof of \Cref{thm:wsl-fold-unfold-complete}.

\section{Rogue Models}
\label{sec:rogue-models}
Imbued with our new understanding of fold/unfold
as the proof mechanism of WSL (and not SL),
we can now identify proof failures of fold/unfold
as stemming from 
the semantic gap between \seplog{} and \wsl{},
exemplified by \emph{rogue, nonstandard} counter-models:
\wsl{} \fp{} structures that refute validity of entailments
but are not \seplog{} \lfp{} structures intended.
Soundness of fold/unfold for \wsl{}
means that these counter-models cannot be eliminated 
by such proof mechanisms,
no matter how many efforts are invested.
The ability to describe, visualize and find these models
allows us to better understand tool failures,
improve diagnosability of these failures,
and provide guidance to the user
in how to proceed with the proof
(e.g., by employing induction rules, which are beyond fold/unfold).
However, precisely defining a nonstandard model could be nontrivial,
as counter-models to entailment in \wsl{} may be infinite
and in some cases, \emph{must} be infinite, as we show next.

In this section we harness the \fo{} encoding 
presented in \Cref{sec:fo-encoding} for the purpose of finding rogue models.
We generalize from the theory-free, quantifier-free formulas,
that we discussed in the previous section,
to all \eprwsl{} formulas the \fo{} encoding supports.
Namely, we observe that beyond a theoretical crutch 
for proving the completeness of \wsl{},
the \fo{} encoding also gives us the ability to use 
existing \fol{} tools 
to show validity of entailments,
in a manner similar to previous work,
and to find infinite counter-models to entailments,
which is unique to this work.

\paragraph{Rogue models}
Formally, a rogue model in this context
is any structure in the class of structures of \wsl{}
that is not part of the class of structures of \seplog{}.
Given a SID $\sys$,
a rogue model of $\sys$ is any structure
$M \in \parens{
	\Models_\wsl \cap \Models_\fp(\sys)
} \setminus \parens{
	\Models_\seplog \cap \Models_\lfp(\sys)
}$.
The key property of the \fo{} encoding
is that it is \emph{model preserving}.
A counter-model $M_\fo$ to the encoded entailment
corresponds to a (unique) heap-structure $M$
that is a \fp{} structure of the system and
satisfies the antecedent of the entailment,
but not the consequent.
Further, $M_\fo$ gives us the only candidate satisfying heaplet
for any sub-formula,
thus resolving the second-order quantification over heaplets,
and allowing us to easily verify 
and understand why $M$ is a counter-model.

\subsection{Heap-Reducing Systems of Inductive Definitions}
\label{sec:heap-reducing}
Though in theory rogue models may be finite or infinite,
we see that in practice systems of inductive definitions
exhibit only infinite nonstandard models.
As discussed in existing literature,
the nature of most real-world systems of inductive definitions
is such,
that any recursive use of a predicate always appears 
under a smaller heap.
If the heap is finite, then
any predicate may only be unfolded a finite number of times
before reaching a \emph{base case}
(i.e., a case with no recursive applications).
Intuitively,
this means that any tuple of elements 
satisfying an inductive predicate
does so by virtue of a finite chain of unfoldings
that reach a base case,
and thus \lfp{} and \fp{} semantics coincide. 

We follow the spirit of~\cite{separation-predicates},
where instead of this intuitive definition of heap-reducing,
\citet{separation-predicates} opt for a stricter,
syntactic definition of heap-reducing.

\begin{definition}
	\label{def:heap-reducing}
	A system of inductive definitions $\sys$
	is \emph{heap reducing}
	if for every inductive predicate $P$,
	each case $\rho^P_j$
	either contains no inductive predicates,
	or there exists some $\varphi$
	s.t. $\rho^P_j$ is equivalent to
	$t \points{\dots} * \varphi$
	(and all applications of inductive predicates
	appear in $\varphi$).
\end{definition}

Note that restricting ourselves to heap-reducing SIDs 
is a mild requirement
compared to stricter requirements like progress
(see, e.g.,~\cite{tree-width-sl}),
used in definitions of decidable fragments of \seplog{}.
Still, this heap-reducing requirement 
is enough to erase the distinction 
between \fp{} and \lfp{} models 
in heap-finite structures
(as in the standard \seplog{} semantics).

\begin{claim}
	\label{thm:heap-reducing-fp-lfp}
	Given a heap-reducing system $\sys$,
	any heap-finite \fp{} structure for $\sys$
	is also a \lfp{} structure for $\sys$:
	$\Models_\seplog \cap \Models_\fp(\sys)
	=
	\Models_\seplog \cap \Models_\lfp(\sys)
	$.
\end{claim}

\Proof{thm:heap-reducing-fp-lfp}{
	For simplicity, we give the proof for the case of a single-predicate system,
	but the same approach applies for the general case.
	
	Let $M$ be a heap-finite, \fp{} structure for a heap-reducing system $\sys$,
	and let us assume towards that $M$ is not a \lfp{} structure for $\sys$.
	In particular, let us consider some predicate $P$,
	elements $d_1, \dots, d_k$ and finite heaplet $\eta$
	s.t.
	$\tuple{d_1, \dots, d_k, \eta} \in P^M$
	but 
	$\tuple{d_1, \dots, d_k, \eta} \notin P_\lfp$,
	where 
	$P_\lfp$ is the least set such that
	$P_\lfp = \sys^M(P_\lfp)$
	(where
	$\sys^M$ is the transformer of the system under $M$,
	as defined in \Cref{def:sid-sat}).
	Let us denote 
	$M' = M\SingSubstitute{P}{P_\lfp}$.
	
	Since $\tuple{d_1, \dots, d_k, \eta} \in P^M$,
	there must exist some $j$
	s.t.
	$M, v, \eta \models \rho^P_j$,
	where
	$v = \SingSubstitute{\Vec x}{\Vec d}$.
	If $\rho^P_j$ is a base-case, then by definition
	$M, v, \eta \models \rho^P_j$
	does not depend on the interpretation of $P$ 
	and thus must also hold for
	$M', v, \eta \models \rho^P_j$
	and thus 
	$\tuple{d_1, \dots, d_k, \eta} \in P_\lfp$ ---
	contradiction.
	
	Otherwise, $\rho^P_j$ must be a recursive, heap-reducing case.
	Any inductive-predicate application $P(t_1, \dots, t_k)$ 
	appearing in $\rho^P_j$ appears so positively,
	and thus if
	$M, v, \eta \models \rho^P_j$ then
	$M, v, \eta_1 \models P(t_1, \dots, t_k)$
	for some heaplet $\eta_1 \subsetneq \eta$.
	
	By the same logic,
	$M, v, \eta_1 \models P(t_1, \dots, t_k)$
	due to some non-base case,
	and thus we get an infinite sequence of heaplets
	$\eta \supsetneq \eta_1 \supsetneq \eta_2 \supsetneq \dots$ ---
	contradiction since $\eta$ is finite.
}

An immediate conclusion of the above claim is
that any rogue model of $\sys$,
that is,
a model for $\sys$ under \wsl{} semantics
but not under \seplog{} semantics,
must have an infinite set of heap locations
and/or be a nonstandard model of the background theory.
Thus, the ability to represent and find infinite rogue models
is \emph{necessary},
as in practice all rogue models are infinite,
when we limit ourselves to the standard model of the background theory.

\begin{conclusion}
	\label{thm:heap-reducing-infinite}
	A rogue model 
	for a heap-reducing SID $\sys$
	has to be in
	$\parens{\Models_\wsl \setminus \Models_\seplog} \cap \Models_\fp(\sys)$,
	and if it extends the standard model of the background theory,
	its location domain must be infinite.
\end{conclusion}

\Proof{thm:heap-reducing-infinite}{
	By definition, a rogue model for SID $\sys$
	has to be in
	\[
	\parens{\Models_\wsl \cap \Models_\fp(\sys)} \setminus \parens{
		\Models_\seplog \cap \Models_\lfp(\sys)
	}.
	\]
	By \Cref{thm:heap-reducing-fp-lfp}, this can be equivalently denoted as
	\[
	\parens{\Models_\wsl \cap \Models_\fp(\sys)} \setminus \parens{
		\Models_\seplog \cap \Models_\fp(\sys)
	},
	\]
	which, by the distributivity of $\cap$ over $\setminus$, is equivalent to
	\[
	\parens{\Models_\wsl \setminus \Models_\seplog} \cap \Models_\fp(\sys)
	.
	\]
}

\subsection{Symbolic Structures}
\label{sec:symbolic-structs}
Rogue counter-models for entailments may be infinite,
and as discussed in the previous subsection,
for the very common case of a heap-reducing system,
any rogue counter-model \emph{must} be infinite.
In order to find and present such models
we first need a way to represent them.

One such representation is \emph{symbolic structures}~\cite{infinite-needle,axe-em},
which were successfully used recently to find models
for formulas in FOL.
Symbolic structures encode infinite structures 
in a finite way,
describing the infinite shape of the structure
symbolically,
with terms and formulas in Linear Integer Arithmetic (LIA).
Though used primarily for uninterpreted formulas 
in~\cite{infinite-needle,axe-em}, symbolic structures can be naturally extended
to support formulas combining uninterpreted logic
with LIA,
which proves useful for us,
as many entailments in WSL use integers as a background theory.

We start by recalling the definition of symbolic structures,
now extended to support integers.
Symbolic structures use finitely many nodes 
to represent the elements
in the domain of an explicit structure,
where each node represents a set of explicit elements
by a copy of some subset of the integers,
specified by a formula in \lia{},
called the bound formula.
The set of nodes form the symbolic domain of the structure
(for each sort).
Our extension forces the domain of sort $\intSort$
to comprise of exactly one node,
representing a full copy of the integers.
Interpretations of functions and relations
in the explicit structure are defined symbolically
over the nodes in the symbolic domain,
with \lia{} terms and formulas 
that relate specific elements within the nodes.

\begin{definition}[Extension of~\cite{infinite-needle}]
	\label{def:sym-struct}
	A \emph{symbolic structure} for a vocabulary
	$\Sigma \uplus \intVocab$
	over sorts $\braces{ \fgSort, \intSort}$
	is a triple
	$S = \parens{\Dom, \Bound, \Interp}$, where
	\begin{itemize}
		\item $\Dom$ is the \emph{domain} of $S$,
		mapping $\fgSort$
		to a \emph{finite, non-empty}
		set of nodes $\Dom(\fgSort)$
		and mapping $\intSort$ to the singleton
		$\Dom(\intSort) = \braces{ n_\intSort }$.
		
		\item $\Bound$ maps each node $n$ to a 
		\emph{bound formula},
		a satisfiable \lia{} formula $\Bound(n) \in \qfLia$
		with at most one free variable $i$,
		with $\Bound(n_\intSort) = \true$.
		
		\item $\Interp$ is the interpretation of symbols in
		$\Sigma \uplus \intVocab$.
		Each constant symbol $c \colon \sort$ 
		is interpreted by a pair
		$\Interp(c) = \tuple{n, z}$
		where $n \in \Dom(\sort)$,
		$\SingSubstitute{i}{z} \models_\lia \Bound(n)$.
		Each function symbol 
		$f \colon \sort_1 \tdots \sort_k \to \sort$
		is interpreted as a function $\Interp(f)$
		that maps nodes 
		$n_1 \in \Dom(\sort_1), \dots, n_k \in \Dom(\sort_k)$
		to a pair of node $n \in \Dom(\sort)$ and \lia{} term $t$
		such that $\FreeVars(t) \subseteq \braces{i_1, \dots, i_k}$
		and
		$\bigland_{j=1}^k \Bound(n_j)\SingSubstitute{i}{i_j}
		\models_\lia \Bound(n)\SingSubstitute{i}{t}
		$.
		A relation symbol $R \colon \sort_1 \tdots \sort_k$
		is interpreted as a function $\Interp(R)$
		that maps nodes
		$n_1 \in \Dom(\sort_1), \dots, n_k \in \Dom(\sort_k)$
		to a \lia{} formula $\varphi \in \qfLia$ such that
		$\FreeVars(\varphi) \subseteq \braces{i_1, \dots, i_k}$.
		Additionally, for the symbols in $\intVocab$
		we require that 
		$\Interp(0) = \tuple{n_\intSort, 0}$,
		$\Interp(1) = \tuple{n_\intSort, 1}$,
		$\Interp(+)(n_\intSort, n_\intSort) = 
			\tuple{n_\intSort, i_1 + i_2}$,
		and
		$\Interp(<)(n_\intSort, n_\intSort) = i_1 < i_2$.
	\end{itemize}
\end{definition}

A symbolic structure should be understood as a finite \emph{representation}
of an explicit, potentially infinite, first-order structure,
denoted by $\Explic S$.
Formally, the explication of a symbolic structure $S$
is a first-order structure 
$M = \Explic S = \parens{\Dom^M, \Interp^M}$
over $\Sigma \uplus \intVocab$
defined as follows.
For each node $n$ in the domain of $S$,
we define its set of explicit elements as
$\Explic n = \braces{
	\tuple{n, z} \mid z \in \Bound(n)
}$,
where we refer to the formula $\Bound(n)$
as the set of integers
$\braces{ z \mid \SingSubstitute{i}{z} \models_\lia \Bound(n) }$.
For each sort $\sort$,
its explicit domain is the set
$\Dom^M(\sort) = \bigcup_{n \in \Dom(\sort)} \Explic n$.
For each constant symbol $c$, $\Interp^M(c)=  \Interp(c)$.
For each function symbol $f$
and nodes $n_1, \dots, n_k$,
let $\Interp(f)(n_1, \dots, n_k) = \tuple{n, t}$,
then for every 
$z_1 \in \Bound(n_1), \dots, z_k \in \Bound(n_k)$,
$\Interp^M(f)\parens{
	\tuple{n_1, z_1},
	\dots,
	\tuple{n_k, z_k}
} = \tuple{
	n,
	\overline{\SingSubstitute{\Vec{i}}{\Vec{z}}}(t)
}$.
For each relation symbol $R \colon \sort_1 \tdots \sort_k$,
then
$\Interp^M(R) = \braces{
	\parens{
		\tuple{n_1, z_1},
		\dots,
		\tuple{n_k, z_k}
	}
	\mid
	n_j \in \Dom(\sort_j),\,
	z_j \in \Bound(n_j)
	\text{ and }
	\SingSubstitute{\Vec{i}}{\Vec{z}} \models_\lia 
	\Interp(R)(n_1, \dots, n_k)
}$.

Unlike the location sort $\fgSort$,
whose domain can consist of an arbitrary number of nodes 
and arbitrary bound formulas, 
we restrict the domain of the integer sort
to have a single node, representing the complete set of integers,
and restrict the interpretations to the standard ones.
This ensures that symbolic structures 
extend the standard model of \intTheory, 
and thus lets us focus on rogue counter-models that arise 
from the other relaxations of \wsl{};
these rogue models must be infinite in the case of heap-reducing SIDs
(\Cref{thm:heap-reducing-infinite}).

\begin{claim}
	\label{thm:sym-struct-int}
	Given a symbolic structure $S$
	over $\Sigma \uplus \intVocab$,
	its explication
	$\Explic S$ 
	extends the standard model of \intTheory, $\intModel$.
\end{claim}

\Proof{thm:sym-struct-int}{
	Let $S$ be a symbolic structure over
	$\Sigma \uplus \intVocab$
	and let $M = \Explic S$ be its explication.
	Let us denote by $M'$ the restriction of $M$
	to symbols from $\intVocab$ and to the domain of
	$\intSort$.
	It is easy to see that the explicit domain of $\intSort$,
	namely,
	$\tuple{n_\intSort, i}$ for any $i \in \Z$,
	together with the interpretations of
	$0, 1, {+}$ and ${<}$
	are isomorphic to the standard model of \lia{}.
}

As shown in~\cite{infinite-needle},
symbolic structures admit decidable model-checking,
by translating any formula $\varphi$
over $\Sigma \uplus \intVocab$
into a pure \lia{} formula $\varphi_\lia$,
such that $\varphi$ is satisfied by a structure $S$
iff $\varphi_\lia$ is valid
(recall that the validity problem of formulas
in \lia{} is decidable).

\cite{infinite-needle} also introduced an efficient way to search
for a satisfying symbolic structure for a given formula.
Through the concept of a \emph{template},
a (possibly infinite) \emph{symbolic search space} of symbolic structures.
A template specifies the general shape of a symbolic structure,
by fixing its domain,
and fixing finite sets of terms and formulas to be used 
in functions, relations and bound formulas.
However, it allows these terms and formulas 
to include free (integer) variables,
thus representing an infinite set of structures.
Given such a template, a formula $\varphi$ is translated into a
\lia{} formula $\varphi_\lia$,
such that $\varphi_\lia$ is satisfiable iff 
any structure within the template satisfies $\varphi$.
Moreover, the satisfying assignment to $\varphi_\lia$
gives a complete definition of the satisfying symbolic structure.

By definition, any entailment that 
is valid in \seplog{} but not in \wsl{}
has a rogue, nonstandard counter-model.
However, 
we are not guaranteed to always find a counter-model 
using the symbolic structures formalism. 
In particular, \Cref{def:sym-struct} only allows 
to represent models 
that extend the standard model of the integers. 
This could possibly be relaxed, 
allowing also for encoding non-standard models of the theory. 
Regardless, we can never hope to find all rogue counter-models 
in this way 
as the class of symbolic structures
is recursively enumerable and model-checking is decidable,
and this would therefore give us 
an RE invalidity procedure for \eprwsl{},
which is not possible since \eprwsl{} has a validity procedure,
but is undecidable (\Cref{thm:wsl-undecidable}).

\neta{excluded distinct-case optimization}

\section{Implementation and Evaluation}
\label{sec:eval}

We implemented the \fo{} encoding described in 
\Cref{sec:fo-encoding}
in a prototype tool
that translates SL entailments modulo SID
into first-order formulas,
where a satisfying model 
for the FO formula
represents a counter-model to entailment
according to the weak semantics.
We discharge these FO formulas both to \zzz{}~\cite{Z3},
for checking unsatisfiability
(thus proving validity)
and to \fest{}~\cite{infinite-needle}
for finding rogue, potentially infinite, nonstandard counter-models.
Our tool is implemented in Python~\cite{python}
and receives as input 
\songbird{}~\cite{songbird} files
that define a system of inductive definitions
and an entailment check, 
given as two SL formulas.
Our tool attempts to both prove validity
(under \wsl{} semantics)
and find a counter-model,
until either attempt is successful
or both time out,
and report the result to the user.
In case of successfully finding a counter-model,
our tool presents the model to the user,
using \fest{}'s syntax for describing symbolic structures.\sharon{mabbe here say that we leave the analysis of counter-models [or the task of synthesizing inductive lemmas based on counter-models] to future work. or is it too early?}

\paragraph{Benchmark suite}
To evaluate the effectiveness 
of the FO translation
in eliciting rogue, nonstandard models of \wsl{}
in real-world problems, we use the \sls{} 
(\songbird{} with Lemma Synthesis) benchmark suite~\cite{sls},
which includes over 700 entailment modulo SID examples,
compiled from existing benchmarks~\cite{slcomp2014,automated-mutual-explicit-induction-sls-benchmark}
and extended with new examples.
The \sls{} benchmark suite contains a variety of inductive definitions
for singly- and doubly-linked lists, nested lists, list segments, 
trees and tree segments,
with nontrivial entailment problems for each kind of system.
About 30\% of the examples use \lia{} as the background theory,
while the rest use no background theories. 
The \sls{} benchmark is made up of 5 parts:
\begin{enumerate*}[(i)]\item \benchmark{sll-valid}, containing examples
	of singly-linked lists, 
	most of which are valid with no need for induction,
	i.e., just by using fold/unfold,
	and thus also valid in \wsl{};
\item \benchmark{slrd-valid}, containing examples 
	for singly-linked, doubly-linked lists and trees,
	where most are valid without induction;
\item \benchmark{slrd-ind}, containing examples
	for singly-linked lists that require induction,
	thus invalid under \wsl{} semantics;
\item \benchmark{slrd-lm} \subbench{w/o integers},
	containing examples for singly-linked lists,
	doubly-linked lists and trees,
	that require inductive lemmas to prove;
	and
\item \benchmark{slrd-lm} \subbench{w/ integers},
	containing examples for singly-linked lists,
	doubly-linked lists and trees 
	that require inductive lemmas,
	where integers are used in formulas and inductive definitions.
\end{enumerate*}
Each part of the benchmark is further subdivided into groups,
where each group defines a SID, 
shared between all examples in that group.
Note that the benchmark does not 
utilize our tool to its full extent,
since it is restricted to the more limited symbolic-heap fragment
that most existing tools are geared at.

\paragraph{Implementation}
Our tool uses a key insight
in order to significantly simplify the FOL queries
sent to \zzz{} and \fest{},
and thus improve success rates:
the overwhelming majority of quantifiers used 
in the benchmark suite are of the form
$\exists \Vec u. \cdots * t \points{\Vec u} * \cdots$
(for some term $t$),
which will be translated on the FO side to
$\exists \Vec u. 
	\cdots \land \parens{\bigland_i m_i(t) = u_i} \land \cdots$.
We observe that we can thus replace every occurrence of $u_i$
with $m_i(t)$
and remove the quantifier entirely.
As each quantifier is transformed into a 
disjunction over all nodes of the symbolic structure,
this optimization allows us to tame the exponential explosion
of formulas produced by \fest{}. 

We build upon \fest{}'s existing library of templates
used for model-finding,
and fine-tune a template that better captures
the general shape of counter-models in separation logic.
The division of the benchmark suite into groups,
where each group shares a SID,
allows us to inspect very few examples manually
to produce this fine-tuned template.

\neta{excluded: distinct case in practice}

\begin{table}
	\begin{footnotesize}
		\caption{Evaluation results.
			We report for each part of the benchmark
			the total number of examples,
			the number of examples that were proved valid 
			through the \fo{} encoding,
			the number of examples where a counter-model was found,
			and the number of timeouts (tool failures) after 30 seconds.
		}
		\label{tbl:eval-results}
		\begin{tabular}{lcrrrr}
			\toprule
			\multicolumn{2}{c}{\bfseries Benchmark}
			& \multicolumn{1}{c}{\bfseries \# Examples}
			& \multicolumn{1}{c}{\bfseries \# Valid}
			& \multicolumn{1}{c}{\bfseries \# Counter-model}
			& \multicolumn{1}{c}{\bfseries \# Timeout}
\\
			\midrule
			\multicolumn{2}{l}{\benchmark{sll-valid}} 
			& 171 & 114 & 4 & 53 \\
			\midrule
			\multicolumn{2}{l}{\benchmark{slrd-valid}} 
			& 151 & 85 & 8 & 58 \\
			\midrule
			\multicolumn{2}{l}{\benchmark{slrd-ind}} 
			& 96 & 0 & 89 & 7  \\
			\midrule
			\multirow{2}{*}{\benchmark{slrd-lm}} 
			& \subbench{w/o integers} & 82 & 5 & 38 & 38 \\
			& \subbench{w/ integers} & 218 & 5 & 26 & 187 \\
			\bottomrule
		\end{tabular}
	\end{footnotesize}
\end{table} 
\paragraph{Evaluation and results}
We run our tool on the SLS benchmark suite using 
a MacBook Pro with an Apple M1 Pro CPU and 32GB RAM,
using
\zzz{} version 4.13.3
and \fest{} version 2.1.1.
The results are summarized in \Cref{tbl:eval-results},
where for each part of the benchmark we report 
how many examples it contains,
how many were proved valid,
for how many our tool produced a counter-model,
and for how many our tool timed out (with a timeout of 30 seconds).
We do not provide comparison with existing tools
since,
to the best of our knowledge,
there are no other tools for finding infinite entailment-violating models
in the context of SL.
For the \benchmark{sll-valid}
and \benchmark{slrd-valid} parts of the benchmark,
which mostly comprise of valid entailments provable without induction,
our tool is able to prove validity for \around{66}\% 
and \around{56}\% of the examples
(respectively)
and find counter-models for a handful of examples.
As the vast majority of examples of these parts 
are valid under \wsl{} semantics,
we note that the failure of our tool is due to insufficient time
for proving the validity of the FOL queries,
rather than inability to find counter-models.
For the remaining parts of the benchmark,
our tool is able to find counter-models for 
\around{93}\% of the examples in \benchmark{slrd-ind},
\around{47}\% in the more complex part
\benchmark{slrd-lm} \subbench{w/o integers},
and \around{12}\% in \benchmark{slrd-lm} \subbench{w/ integers},
which uses \intTheory{} as the background theory,
thus requiring the largest and most complicated symbolic structures
in counter-models.
We conjecture that by examining a few representative
examples from the unsolved problems,
the template used can be further tuned,
and success rates improved.

\paragraph{Partial taxonomy of rogue models}
	We now give a partial taxonomy 
	of the kinds of rogue models that 
	crop up in practical
	\seplog{} entailments with inductive definitions.
	By examining the counter-models produced by our tool
	we identify repeating patterns and archetypes 
	of rogues models.
	In essence,
	these archetypes correspond to 
	the \emph{heap shape} of the rogue model,
	given by the symbolic domain and interpretation of the points-to functions,
	with the specifics of each counter-model
	given by the symbolic interpretation of constants and predicates.
	We describe 6 such archetypes:
	\begin{enumerate}[leftmargin=1.75em, labelsep=0.5em]
		\item 
		\label{archetype:list-1}
		Linked list that never reaches \Null:
		these are made up of two nodes $n_0, n_1$ in the symbolic domain,
		$n_0$ representing \Null{} and $n_1$ representing
		infinitely many locations that form a linked list.
		\neta{cut:where no location points to \Null.}
		This archetype appears mostly in examples with unary
		inductive predicates over lists.
		
		\item 
		\label{archetype:list-2}
		Linked list that never reaches some location other than \Null:
		made up of three nodes $n_0, n_1, n_2$,
		where $n_0$ is \Null{}, $n_1$ is the unreachable location, 
		and $n_2$ represents all other locations which form an infinite linked list.
		This archetype appears in examples with 
		`list-segment' predicates.
		\neta{cut:a binary `list-segment' 
		of some kind.}
		
		\item 
		\label{archetype:list-3}
		Two disjoint, infinite linked lists that never reach \Null:
		made up of three nodes $n_0, n_1, n_2$
		as in archetype \ref{archetype:list-2},
		but with $n_1$ representing an infinite linked-list of locations.
		This archetype appears in more complex `list-segment' examples
		and doubly-linked lists.
		
		\item Tree where one path never reaches \Null:
		similarly to archetype \ref{archetype:list-1}, these are made up of two nodes $n_0, n_1$,
		with $n_0$ representing \Null, and $n_1$ 
		represents infinitely many locations on a branch in the tree, 
		where each points to \Null{} in one branch, 
		but continues the infinite tree in the other.
		This archetype appears in
		examples with inductively-defined trees
		and nested linked-lists.
		
		\item Tree where one path never reaches some location other than \Null:
		made up of three nodes, akin to archetype \ref{archetype:list-2},
		appears in examples with `tree-segments'
		or simple overlaid data-structures.
		
		\item Trees where one path never reaches some location that never reaches \Null:
		akin to archetype \ref{archetype:list-3},
		appears in more complex examples of tree segments and overlaid data-structures.
	\end{enumerate}

\section{Related Work and Conclusion}
\label{sec:relwork}

There are several approaches to automating reasoning for SL, 
including the development of decidable fragments~\cite{Smallfoot,BerdineCalcagnoOHearn2004,BerdineCalcagnoOHearn2005,CookHaaseChristoph2011,PerezRybalchenko2013,PiskacWiesZufferey2013,guarded-wand-pagel2020,twbcade13,EchenimetalPCEGeneralization2021}, search techniques over proof systems~\cite{ChinDavidNguyen2007,ChinDavidNgyuenQin2012,songbird,PerezAntonioRybalchenko2011,ChuJaffarTrinh2015,brotherston11}, and translations to FOL to utilize SMT solvers and FO theorem provers~\cite{ChinDavidNguyen2007,ChinDavidNgyuenQin2012,pek14,PiskacWiesZufferey2013,PiskacWiesZufferey2014,PiskacWiesZufferey2014Tool,madhusudan12,qiu13,focomplete-heap-logics,vipertool,viper-vcgen-technique}. 
Our work is distinguished in this regard since it studies a rich  
logic
(in particular, it is undecidable)
with a general fragment which admits a model-preserving translation into FOL. 

Fold/unfold mechanisms are extremely commonplace, with uses
in logics beyond separation
logic~\cite{boyer88,kaufmann97,kaufmann00,madhusudan12,nguyen08,pek14,qiu13,suter10,amin2014computing,leino2013verifiedcalculations,acl2sedan,passmore20,Vazou18,liquid,SuterKoksalKuncak,BlancKuncakKneussSuter,loding18,fluid23,focomplete-heap-logics,ChinDavidNguyen2007,ChinDavidNgyuenQin2012,brotherston11,leino12,jacobs11}.
These are a \emph{class} of mechanisms as opposed to a
single authoritative algorithm. 
Our work on the
(in-)completeness of these mechanisms for SL with inductive
definitions is inspired by recent
works~\cite{loding18,fluid23} which study the corresponding
question for first-order logics. However, to our knowledge
ours is the first work to develop such foundations for
separation logic. This is nontrivial, since there are many
more sources of incompleteness that interact in complex ways
\ifextended
(see~\Cref{sec:background}).
\else
(see details in the extended version
of the paper~\cite{wheat-chaff-extended}).
\fi
\sharon{revised:}Related to our investigation of incompleteness of fold/unfold mechanisms is 
the work in~\cite{vipercav24}, which studies incompleteness empirically for a range of automated program verifiers that work with separation logic annotations.

There is also a long tradition of studying the completeness of reasoning heuristics in the context of first-order resolution~\cite{SetOfSupport1965,SetOfSupportImproved2021,stickel-theory-resolution,instgen,Ge-deMoura2009CompleteInstantiationSMT,Ihlemann2008LocalTheoryExtensions}.

Definitions with determined heaplets have been studied in early work on separation logics as \emph{precise predicates}~\cite{ohearn04,BerdineCalcagnoOHearn2004,precise-predicates-concurrent,ohearn04concurrency}. There are also a plethora of works that define explicit unique heaplets for formulas in heap logics~\cite{qiu13,framelogictoplas2023,separation-predicates,focomplete-heap-logics}.

Our technique for synthesizing rogue models is based on the recent work on symbolic structures~\cite{infinite-needle}. However, there are other frameworks such as the work in~\cite{BlanchetteLPAR2010} which studies rogue models for first-order logic formulas over algebraic datatypes generated by Isabelle/HOL. Symbolic structures are also reminiscent of abstract domains for representing unbounded structures~\cite{HeapGraphs,tvla-with-rechability}, but we are not aware of a formal connection between the two techniques.

We end this section with a note on inductive reasoning which can be used to overcome fold/unfold proof failures. 
\neta{rephrased from: There are several techniques for automating induction reasoning in separation logic including deductive instantiations of induction axioms, lemma synthesis, and cyclic proofs~\cite{sls,ChuJaffarTrinh2015,EneaSighireanuLemmaSynthesisSL,brotherston11}.}
There are several techniques for automating induction reasoning in separation logic, 
including deductive instantiations of induction axioms, lemma synthesis, and cyclic proofs~\cite{sls,ChuJaffarTrinh2015,EneaSighireanuLemmaSynthesisSL,brotherston11}. Using
the symbolic representation of rogue models found by our technique
to automatically deduce and generate missing lemmas
for proving validity of 
entailments in separation logic is an exciting direction that we leave to future work.

\sharon{merged from conclusion:}
In conclusion, we consider our work to be a first step towards a better understanding 
of \seplog{} tools and, particularly, their failures.
We believe that \wsl{} provides a way to contextualize 
and analyze tool failures,
both on a theoretical level and as a practical framework, which we aim to pursue further in the future.  
While we have focused on fold/unfold reasoning, analyzing reasoning of proof mechanisms beyond 
fold/unfold is also of great interest and left for future work.

\section*{Data-Availability Statement}
An artifact for reproducing the results is available
at~\citep{wheat-chaff-artifact-v3},
and the latest version of the tool is available
at~\citep{wheat-chaff-artifact-latest}. 

\begin{acks}
We thank the anonymous reviewers 
for comments which improved the paper.
The research leading to these results 
has received funding from the
European Research Council 
under the European Union's Horizon 2020 research and
innovation programme (grant agreement No [759102-SVIS]).
This research was partially supported by 
the Israeli Science Foundation (ISF) grant No.\ 2117/23.

\end{acks}
 \bibliography{includes/references.bib}

\ifextended
\newpage
\appendix
\crefalias{section}{appendix}
\section{Background: Sources of Incompleteness in Separation Logic}
\label{sec:background}

In the introduction we described the idea of a \emph{logical characterization} of the completeness of proof mechanisms. To pursue this technique, we must systematically identify and eliminate the sources of incompleteness in standard Separation Logic while preserving the soundness of the proof mechanisms under study. The development of a theoretical framework for defining a class of proof mechanisms and the proof of their soundness with respect to these `ablated' logics is nontrivial and is one of the main contributions of this work.

Although 
the focus of this paper is on the (in-)completeness of fold/unfold mechanisms, in this appendix we focus on presenting a background on separation logic and its various sources of incompleteness. 

\mypara{Ablating Sources of Incompleteness: What does it mean?} We first describe what ablating a source of incompleteness means. In the sequel we will argue that certain design decisions in the definition of separation logic semantics are so powerful that they cause validity checking to be non recursively enumerable. More technically, one can use these design decisions to construct entailments whose validity can capture the non-halting of Turing Machines. Recall that this means there cannot be any sound proof system or proof mechanism that can prove all valid entailments. We also argued in the introduction that this lack of a fixed standard for automation manifests as a practical limitation on the predictability of automation techniques for Separation Logic.

Ablating a particular source of incompleteness then involves defining an alternate Separation Logic whose syntax is identical to that of SL but whose semantics removes or modifies the particular design decision. Consider for example the fact that the semantics of standard SL is defined over finite heaplets. A formula is valid in standard SL if it holds on all finite heaplets. Ablating this design decision yields a Separation Logic where formulas must be evaluated over both finite and infinite heaplets, and a formula is valid only if it holds on all heaplets, finite or infinite. The resulting logic is not identical to standard SL: consider an inductively defined relation that holds if a location is reachable from a designated location $x$. Over finite heaplets, it is true that the set of reachable locations from $x$ has an element whose distance from $x$ 
is larger than the distance of any other element in the set, but this is clearly not true over infinite heaplets, because we can construct an infinite chain of locations reachable from $x$ with ever increasing distances. 

Systematically ablating the various sources of incompleteness and studying the differences between these variants is the key to understanding the connection between the sources of incompleteness and the failure of proof mechanisms.

\subsection*{Systematically Ablating Sources of Incompleteness}

In the remainder of this section we provide some background on the various sources of incompleteness in SL via a series of results known in the literature. We do not present these results formally as they are not the contributions of this work and are intended to serve as an illustration of our own journey towards the ideas developed in this paper.

We first remove the magic wand from our investigation, as it is one of the most well-known sources of incompleteness in separation logic. It in fact  makes SL equivalent to Second-Order Logic~\cite{magic-wand-complexity}. The hardness of reasoning with magic wands is also evident from the fact that many automated tools do not support it~\cite{slcomp}. We therefore restrict ourselves to only considering Separation Logic formulas without magic wands. 
In the sequel we use SL to refer to separation logic \emph{without magic wands}.

\mypara{Source 1: Finite Heaplets} The definition of semantics in standard SL considers satisfaction only over finite heaplets. The consequent definition of validity of entailments to only consider finite heaplets is a key source of incompleteness in SL. Reasoning about validity over finite models is extremely hard; the corresponding result for First-Order Logic is Trakhtenbrot's theorem:

\begin{proposition}[Trakhtenbrot's Theorem]
    Validity of first-order logic over finite models is not recursively enumerable.
\end{proposition}

The above proposition can be proven by constructing linked lists that encode partial runs of Turing Machines that do not encounter a halting state. Since the formulas are only evaluated over finite models, the resulting predicate is valid if and only if there is truly no finite amount of time where the run of the Turing Machine does not encounter a halting state. It is easy to see that we can also express this construction in Separation Logic, which yields the following claim:

\begin{claim}
\label{clm:finite-heaplets-incompleteness}
    Finite heaplets are a source of incompleteness in SL (even without magic wands). \end{claim}

We can ablate this source of incompleteness by considering validity over both finite and infinite heaplets. Note that this is a weakening (or relaxation) of SL, in the sense that validity over the ablated logic implies validity in SL, but not the other way around.

\mypara{Source 2: Inductive Definitions with Least Fixpoint Semantics} Inductively defined predicates in SL are interpreted as the least fixpoint of their definitions. This is another well-known source of incompleteness in the literature with a fairly obvious argument: one can define natural numbers by associating them with linked lists of the corresponding length, and define addition and multiplication as predicates that take two linked lists and determine whether their lengths add up (resp. multiply) to the length of a third linked list. This then yield's incompleteness by G\"{o}del's incompleteness theorem for arithmetic.

\begin{claim}
    Inductive definitions with least fixpoint semantics are a source of incompleteness in SL. \end{claim}

We can ablate this source of incompleteness by considering interpretations of defined predicates that are consistent with \emph{any} fixpoint as opposed to the least one. This clearly makes a difference: if we say that a predicate $R$ holds on $x$ if and only if it holds on $\mathit{next}(x)$, then on a cycle the least fixpoint would mark $R$ to be false on all the elements, whereas marking them all true is clearly consistent with the recursion and is a valid fixpoint.

The consideration of arbitrary fixpoints is also a weakening of SL. Validity in the ablated logic implies validity in SL, but not the other way around.

\mypara{Source 3: Combination of Background Sorts with Intended Models} SL formulas written over a combination of sorts are given semantics by interpreting operations over background sorts, say integers, over a \emph{standard} or intended models of the sort. For integers this is the usual number line with the usual definition of arithmetic operators. However, even though individual background sorts can admit r.e. procedures for validity, combining them can result in incompleteness. This is also well known: consider for example the combination of (a) integers with addition and (b) uninterpreted functions. Individually, validity in these logics is recursively enumerable. The former is Presburger Arithmetic and the latter is G\"{o}del's completeness theorem for FOL. However, when the two are combined, we can write a formula for an uninterpreted predicate $\mathit{mul}$ such that $\mathit{mul}$ behaves exactly like multiplication over the integers. This then results in incompleteness. 

\begin{claim}
    Combinations of Background Sorts with Intended Models is a source of incompleteness in SL. \end{claim}

Ablating this source of incompleteness is rather tricky. It turns out that automated reasoning techniques for logics utilize certain axiomatizations for background theories. Therefore, it is reasonable to suggest that one can consider models of SL with background sorts where the sorts are not only realized by the intended model, but any model consistent with their axiomatized theory. This is also clearly a weakening of SL since it expands the set of models considered. Validity in the ablation implies validity in SL.

\mypara{Ablations: To Combine or Not to Combine?} In the above discussion, we identified several sources of incompleteness and suggested ablations for them. Is it possible that ablating only one of the sources or a strict subset of them is sufficient to recover a well-behaved logic? We turn once again to completeness to answer this question. If our goal is to identify a weakening of SL for which fold/unfold mechanisms are complete, then surely it must be complete in the first place! We argue below that a combination of all the above ablations is needed, because imposing fewer ablations also yields incomplete logics:

\begin{proposition}
SL with finite heaplets ablated, i.e., interpreted over both finite and infinite heaps is incomplete.  
\end{proposition}

When we only ablate finite heaplets we still have least fixpoint inductive definitions. We can then define linked lists that encode partial non-halting runs of a Turing Machine as a recursive definition. This effectively reduces the consideration to finite linked lists since least fixpoint semantics would not mark any infinite chains or cycles as valid linked lists in the first place. Therefore we can say that this predicate holds if and only if there are no halting runs of the Turing Machine, which yields incompleteness. 

Observe that the above argument is agnostic to the presence of background sorts. Therefore, even ablating background sort models to all models of the theory would still admit incompleteness:

\begin{proposition}
SL with both finite heaplets and intended models for background sorts ablated is incomplete.
\end{proposition}

We turn to ablating least fixpoint semantics for recursive definitions. Observe that if we continue to have finite heaplets, then we can define the non-halting partial runs all the same. In fact, we show in Section~\ref{sec:heap-reducing} (see Theorem~\ref{thm:heap-reducing-fp-lfp}) that for a class of definitions that includes linked lists, least fixpoints are the only fixpoints when considering finite heaplets. This yields the following result:

\begin{proposition}
SL with least fixpoint semantics for definitions ablated is incomplete.
\end{proposition}

Again, the above construction does not use background sorts at all, therefore:

\begin{proposition}
SL with both least fixpoints and intended models for background sorts is incomplete.
\end{proposition}

This then suggests that we ought to ablate both finite heaplets and least fixpoints, and perhaps intended models for the background as well. This turns out to be insufficient! We first state the following result due to~\cite{focomplete-heap-logics} before discussing the ramifications:

\begin{proposition}[Result due to~\cite{focomplete-heap-logics}]
SL with both finite heaplets and least fixpoint semantics for definitions ablated is incomplete. This holds even without background sorts, and therefore adding the ablation intended models for background sorts is still incomplete.
\end{proposition}

This is incredibly surprising given prior works on logical characterizations of completeness for first-order logics~\cite{loding18,fluid23}. For first order logics, it turns out to be sufficient to consider semantics over infinite heaplets, with arbitrary fixpoint interpretations for recursive definitions, and models of background theory axiomatizations to yield a complete logic. The resulting logic is in fact FOL. However, in our case there happens to be \emph{yet another source of incompleteness}.

\mypara{Source 4: Second Order Quantification over Witnessing Heaplets} Consider the entailment $\alpha \models \beta * \gamma$. Let us take the view of proving validity by refutation. In order to show the above entailment valid, we would try to come up with a model of $\alpha$ such that $\beta * \gamma$ does not hold. If we exhaustively cover all possibilities and show that in each case such a model is not possible, then the entailment is valid. However, observe that to demonstrate that a particular model $s,h$ that satisfies $\alpha$ \underline{does not} satisfy $\beta * \gamma$, we have to argue that there is no way to split the domain of $h$ into disjoint parts such that one of them witnesses $\beta$ and the other witnesses $\gamma$. This is second order quantification! The scenario is further complicated by the fact that the right-hand side of the entailment could be an inductive defined relation with a $*$ in its definition, so there is an explosion of witnessing heaplets to cover in order to ascertain whether the entailment can be refuted. This gadget is so powerful that it can be used to show incompleteness~\cite{focomplete-heap-logics}.

The observation that SL semantics imposes an implicit second-order
quantification is not new. Early work on separation logic sought to avoid this
issue by defining a class of precise predicates, which are inductively defined
predicates such that ``for any given store and heap, there is at most one
subheap that satisfies it; and so every predicate cuts out an unambiguous area
of storage''. In our work we similarly avoid this source of incompleteness by
defining a fragment of separation logic formulas with \emph{determined heaps}.

We conclude this section with the following theorem, which is the first main contribution of this paper:

\begin{theorem}
    SL without (a) magic wands, (b) finite heaplets, (c) least fixpoint semantics for inductive definitions, for a fragment of formulas that admits (d) determined heaps, admits complete validity checking for entailments.
\end{theorem}

\section{Proofs}
\label{sec:proofs}

\Proofs{}
 \fi

\end{document}